\documentclass[aps,prb,twocolumn,letterpaper,showpacs,citeautoscript]{revtex4-1}

\usepackage{amsmath,amssymb}
\usepackage{bm}
\usepackage{graphicx}
\usepackage{epstopdf}
\usepackage{latexsym}
\usepackage{subfigure}
\usepackage{color}
\usepackage{dsfont} 
\usepackage{wasysym} 
\usepackage[T1]{fontenc}
\usepackage{ae,aecompl}

\newcommand{\ra}{\rangle}
\newcommand{\la}{\langle}
\newcommand{\curF}{{\cal F}}

\newcommand{\ve}{\vert}

\newcommand{\ibq}{{\bf{ q}}}
\newcommand{\ibr}{{\bf{ r}}}
\newcommand{\ibQ}{{\bf{ Q}}}
\newcommand{\vphi}{\varphi}
\newcommand{\bk}{{\bf k}}
\newcommand{\curN}{{\cal N}}
\newcommand{\bse}{\begin{subequations}}
\newcommand{\ese}{\end{subequations}}
\newcommand{\uxy}{u_{xy}}
\newcommand{\hepsilon}{{\hat \epsilon}}
\newcommand{\bphi}{\boldsymbol{\phi}}
\def\rf#1{(\ref{#1})}

\begin{document}

\title{Magnetic orders, excitations, and phase transitions in Fe$_{1+y}$Te}

\author{G. Chen}
\affiliation{Department of Physics, University of Colorado, Boulder, CO, 80309-0390, U.S.A.}
\author{S. Choi}
\affiliation{Department of Physics, University of Colorado, Boulder, CO, 80309-0390, U.S.A.}
\author{L. Radzihovsky}
\affiliation{Department of Physics, University of Colorado, Boulder, CO, 80309-0390, U.S.A.}

\begin{abstract}
  We study the magnetic properties of Fe$_{1+y}$Te, a parent compound
  of the iron-based high-temperature superconductors. Motivated by
  recent neutron scattering experiments, we show that a spin $S=1$
  exchange model, supplemented by a single-ion spin anisotropy,
  accounts well for the experimentally observed low temperature
  magnetic phase diagram, that exhibits a commensurate bicollinear
  order at low Fe dopings ($y \lesssim 0.12$) and an incommensurate
  spin-spiral order at high Fe dopings ($y \gtrsim 0.12$).  We suggest
  that the commensurate-incommensurate transition at $y \simeq 0.12$
  is due to the competition between the exchange interaction and the
  local spin anisotropy.  At low Fe dopings, the single-ion spin
  anisotropy is strong and pins the spins along the easy axis, which,
  together with the spatially anisotropic exchanges, induces a unusual
  bicollinear commensurate magnetic order. The low-energy spin-wave
  excitation is gapped due to the explict breaking of spin-rotational
  symmetry by the local spin anisotropy. At high Fe dopings, the
  single-ion anisotropy is weak, and the exchange favors an
  incommensurate coplanar state.  The incommensurate magnetic
  wavevector averages out the spin anisotropy so that a gapless
  low-energy spin-wave excitation is obtained.  We also analyze the
  low-energy hydrodynamic model and use it to describe the
  magneto-structural transition and the static and dynamical spin
  structure factors across the magnetic ordering transitions.
\end{abstract}

\date{\today}


\maketitle

\section{Introduction}
\label{sec:sec1}

Initiated by Hosono and co-worker's discovery of iron-based
high-temperature superconductivity in fluorine-doped
LaOFeAs\cite{Kamihara08}, there have been tremendous research
activities and developments in the area of iron-based superconductors.
While searching for higher transition temperature and its mechanism,
many classes of materials were discovered and analyzed extensively,
theoretically and experimentally\cite{Paglione10}. The most well-known
materials are the 1111 (e.g. LaOFeAs) and 122 (e.g. BaFe$_2$As$_2$)
compounds, conventionally referred to as iron arsenide materials.

These iron-based superconductor exhibit many interesting features that
have attracted considerable attention\cite{Paglione10}.  Their phase
diagram exhibits some similarities to that of cuprate superconductors,
with the pairing mechanism that is believed to be unconventional
(i.e., non-phonon mediated).  Concomitant with this is the
superconducting order parameter that is predicted to be (with some
experimental evidence~\cite{Grafe08,Chen10}) of an unconventional,
extended ($s_{++}$ and $s_{\pm}$, alternating sign around the
Brillouin zone but fully gapped on the Fermi pockets) $s$-wave type.
Angle resolved photoemission spectroscopy (ARPES) and inelastic
neutron scattering studies suggest that the Fermi surface nesting
accompanied by the spin density wave plays a central role for
mediating the superconducting mechanism.

More recently discovered, the so-called 11 materials (e.g. FeSe and
FeTe based compounds) also show superconductivity with doping of
sulfur or selenium.  Their simpler structure, with no atoms at the
interplanar layer, is hoped to be present a simpler challenge of
uncovering the nature of the pairing mechanism, but still to shed
light more generally on iron-based and other strongly correlated
superconductors.

\begin{figure}[t]
\includegraphics[width=7cm]{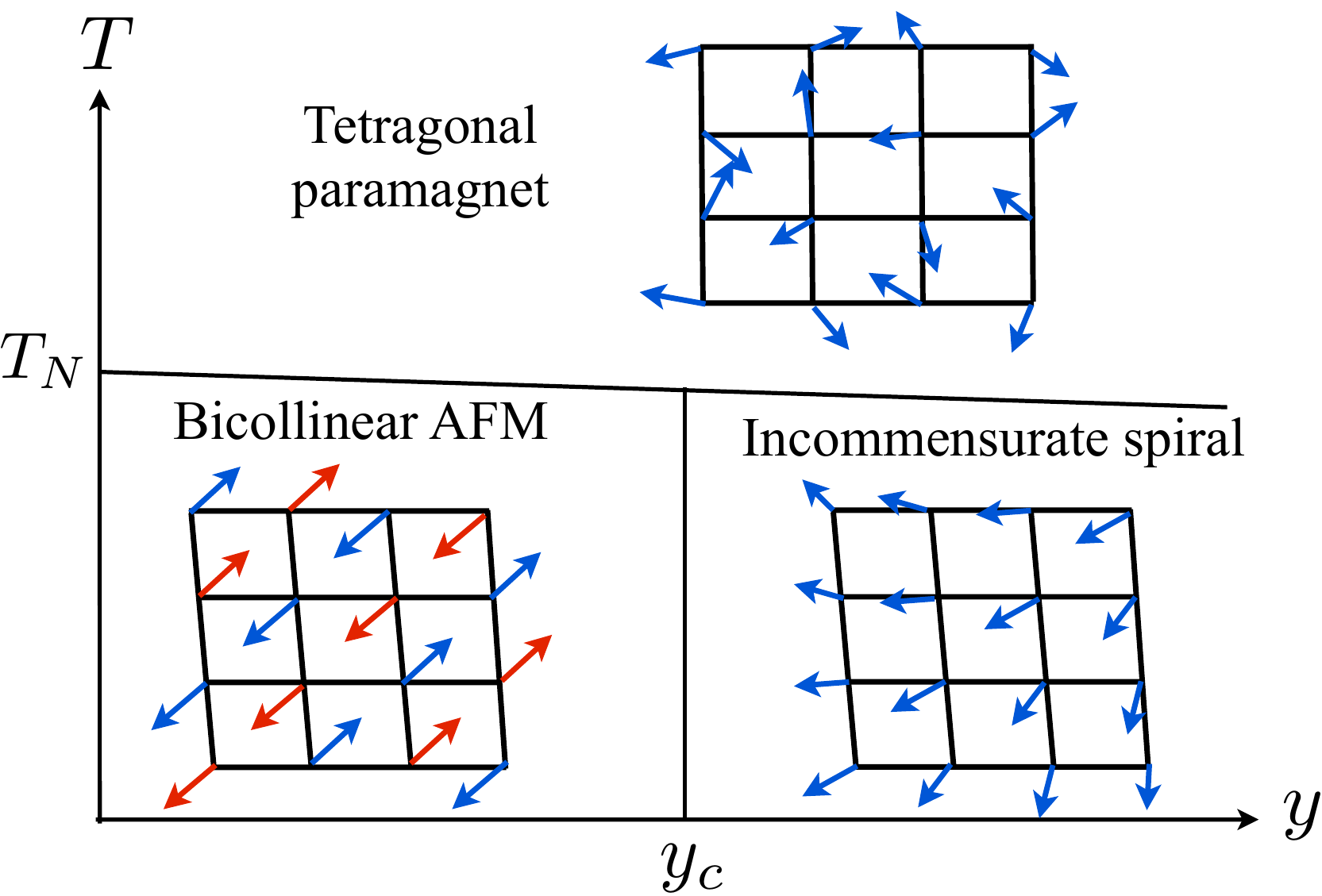}
\caption{The schematic temperature ($T$) Fe-doping ($y$) phase diagram
  for Fe$_{1+y}$Te. $y_c \simeq 0.12$ and the critical temperature
  $T_N\simeq 65K$. In the magnetic ordered phase, the system has an
  orthorhombic crystal structure, and the incommensurate spiral is in the 
  $bc$ plane. }
\label{fig1}
\end{figure}

With magnetism believed to be central to high temperature
superconductivity and interesting in its own right, much attention has
also recently turned to the magnetic parent compounds, such as the
self-doped
Fe$_{1+y}$Te\cite{Bao09,Li09,Subedi08,Xia09,Turner09,Fang09}.
Fe$_{1+y}$Te is observed to exhibit a number of novel characteristics
(see Fig.~\ref{fig1}).  The most interesting of these is a unusual
bicollinear antiferromagnetic (AFM) state, with a commensurate, planar
spin-spiral order characterized by $[\pi/2, -\pi/2]$ ordering
wavevector.  A first-order thermal transition to this state is
accompanied by a structural transition to an orthorhombic state (with
a slight monoclinicity) in low Fe doped samples\cite{Bao09}.  At low
temperatures the magnetic order also undergoes a commensurate to
incommensurate transition with a critical iron doping at $y_c \simeq
0.12$, with low doping corresponding to the commensurate
phase\cite{Li09}.

Measurement of the spin susceptibility with a large magnetic moment of
order $2\mu_B$ (at 5K in a $y=0.068$ sample\cite{Li09}) and no Fermi
surface nesting observed in DFT \cite{Subedi08} and ARPES \cite{Xia09}
suggest that (despite its metallic nature) a local moment description
may be sufficient to capture magnetism in Fe$_{1+y}$Te compounds.
This is supported by first-principles electron structure calculations
that observe the formation of the iron local
moments\cite{Zhang09,Ma09}.  There have also been several theoretical
studies based on the local moment
description\cite{Turner09,Fang09,Yin10,Zhang09,Ma09}.  Turner {\it et
  al.} assumed that the electrons are localized and the structural
transition is driven by an orbital ordering resulting from Jahn-Teller
coupling\cite{Turner09}.  Their model consists of antiferromagnetic
superexchange and ferromagnetic double exchange, which together favor
an incommensurate (close to bicollinear) state, as well as a
biquadratic exchange which can then drive the system to the
commensurate bicollinear AFM state.  Fang {\it et al.} also developed a
local spin model with a rather complicated exchange
interaction\cite{Fang09}.  They obtained a rich phase diagram that
includes the two relevant phases observed in Fe$_{1+y}$Te. Yin {\it et
  al.} unified the two pictures based on itinerant electrons and
localized moments in Fe$_{1+y}$Te in analogy with
manganites\cite{Yin10, Sen07}.  They pointed out the sensitive
competition between the superexchange and orbital-degenerate
double-exchange ferromagnetism, finding several collinear states
including the bicollinear AFM state.  Although these models are
claimed to obtain the bicollinear AFM and incommensurate spin-spiral
states observed in Fe$_{1+y}$Te, the underlying spin-rotational
symmetry in the spiral plane of these models predicts a gapless spin wave excitation,
which is inconsistent with experimental observations of a gapped
spectrum in the commensurate bicollinear AFM state for $y \lesssim
0.12$\cite{Stock11,Lipscombe11}.

The main experimental motivation for our theoretical study is the
recent neutron scattering measurements on the Fe$_{1+y}$Te samples
with a series of Fe dopings\cite{Bao09,Stock11,Rodriguez11}.  Besides
being consistent with the schematic phase diagram in Fig.~\ref{fig1},
the experimental findings provide additional information about the
properties of different magnetic phases. In particular, in Parshall
{\it et al.}'s experiment on a $y=0.08$ sample, it has been observed
that an {\it incommensurate} inelastic peak (at the wavevector
$[0.45\pi,-0.45\pi]$) in the dynamical spin structure factor
precipitously shifts to a commensurate position (at the wavevector
$[\pi/2, -\pi/2]$)\cite{Parshall12}.  Moreover, a gapped spin-wave
spectrum is observed in the dynamical spin structure factor for the
samples with the bicollinear AFM order, while a gapless spin-wave spectrum
is obtained for the samples with the incommensurate order.

To account for the previous and current experiments, we take the local
moment description and consider a microscopic spin model for
Fe$_{1+y}$Te in Sec.~\ref{sec2}.  Clearly, a gapped spin-wave spectrum
for the bicollinear AFM state suggests the explicit breaking of the spin
rotational symmetry at the Hamiltonian level. Such a spin anisotropy
is certainly allowed on symmetry grounds by the orthorhombic crystal
structure in the ordered phase.  Therefore, we introduce a {\it
  single-site spin anisotropy} for the localized $S=1$ spin moment.
In addition, we choose the spin exchange introduced by Turner {\it et
  al.} but abandon the superfluous biquadratic exchange. The
single-site spin anisotropy, together with the spin exchange,
naturally generates the schematic phase diagram in Fig.~\ref{fig1}.
For the bicollinear AFM state at low Fe doping, the lattice distortion is
large and the spin anisotropy is expected to be strong. The strong
spin anisotropy favors collinear spin states.  The spin exchange
further selects the bicollinear AFM state observed in the experiment.  The
spin-wave spectrum of this state is found to be gapped. As the Fe
doping is increased, the lattice distortion is reduced and the spin
anisotropy is also expected to be reduced.  In the end, the dominant
spin exchange favors an incommensurate state.  This incommensurate
spin state averages out the local spin anisotropy, which effectively
restores the spin-rotation symmetry and leads to a gapless spin-wave
spectrum. We also provide a possible microscopic origin for the
single-ion spin anisotropy by introducing the magnetoelastic coupling
in Sec~\ref{sec2e}. 

In Sec.~\ref{sec3}, we focus on a phenomenological continuum Landau theory
and provide a mean-field analysis of the magnetostructural
transition in Fe$_{1+y}$Te. We obtain a global phase diagram that
includes both paramagnetic and AFM phases with 
tetragonal or orthorhombic structures. 
In Sec.~\ref{sec4}, we then discuss the low temperature AFM phases of orthorhombic phase. 
With a proper parameterization of the magnetic order, we map the continuum 
theory to a sine-Gordon model and explain and analyze the 
commensurate-incommensurate transition from the bicollinear AFM state to the 
incommensurate AFM state upon Fe dopings, calculating the static spin structure 
factor near the transition.

In Sec.~\ref{sec5}, utilizing this continuum model of Sec.~\ref{sec4}
and general conservation and symmetry principles we propose a
hydrodynamic theory (extensions of model E and F in the
Halperin-Hohenberg classification\cite{Halperin69,Hohenberg77}) for a finite
temperature, low frequency, long wavelength description of the
system. We use it to calculate the static and dynamical spin structure
factors in different paramagnetic and magnetically ordered phases.

The rest of the paper is organized as follows. In Sec.~\ref{sec2}, we
introduce the microscopic model that is an extension of the model in
Ref.~\onlinecite{Turner09}, incorporating a key new ingredient of
single-ion spin anisotropy and omitting the superfluous biquadratic
exchange. We use it to obtain the classical phase diagram and the
low-energy magnetic excitation spectrum.  We study
the magnetostructural transition 
in Sec.~\ref{sec3} and discuss the commensurate-incommensurate transition 
In Sec.~\ref{sec4}. In Sec.~\ref{sec5}, we compute
the static and dynamical spin structure factors in the different
magnetic phases.  We conclude the paper in Sec.~\ref{sec6} with a
summary and discussion of our predictions.

\section{Model of $\text{FeTe}$}
\label{sec2}

\subsection{Microscopic lattice model}
\label{sec2a}

Our microscopic model for Fe$_{1+y}$Te is based on the lattice $S=1$
exchange model introduced by Turner {\it et al}~\cite{Turner09}.
Although Fe$_{1+y}$Te compounds are metallic, we assume that the
magnetism in Fe$_{1+y}$Te is described by localized $S=1$ spin
moments.  The main difference between our model and that of
Ref.~\onlinecite{Turner09} is our introduction of the single-ion spin
anisotropy that explicitly breaks spin-rotational symmetry, which is
expected from the orthorhombic low-temperature crystal structure and
the observed gapped spin-wave spectrum in the bicollinear AFM state in low
Fe doped samples.  Microscopically, the transition to the orthorhombic
state can be argued to be associated with the orbital ordering via
Jahn-Teller coupling~\cite{Turner09}.  However, we will capture it
more simply, phenomenologically through a magnetoelastic coupling,
that we analyze in Sec.~\ref{sec3}. Also in contrast to
Ref.~\onlinecite{Turner09}, we neglect the biquadratic exchange, that
in our view is not necessary to capture the Fe$_{1+y}$Te
phenomenology.

\begin{figure}[t]
\includegraphics[width=8cm]{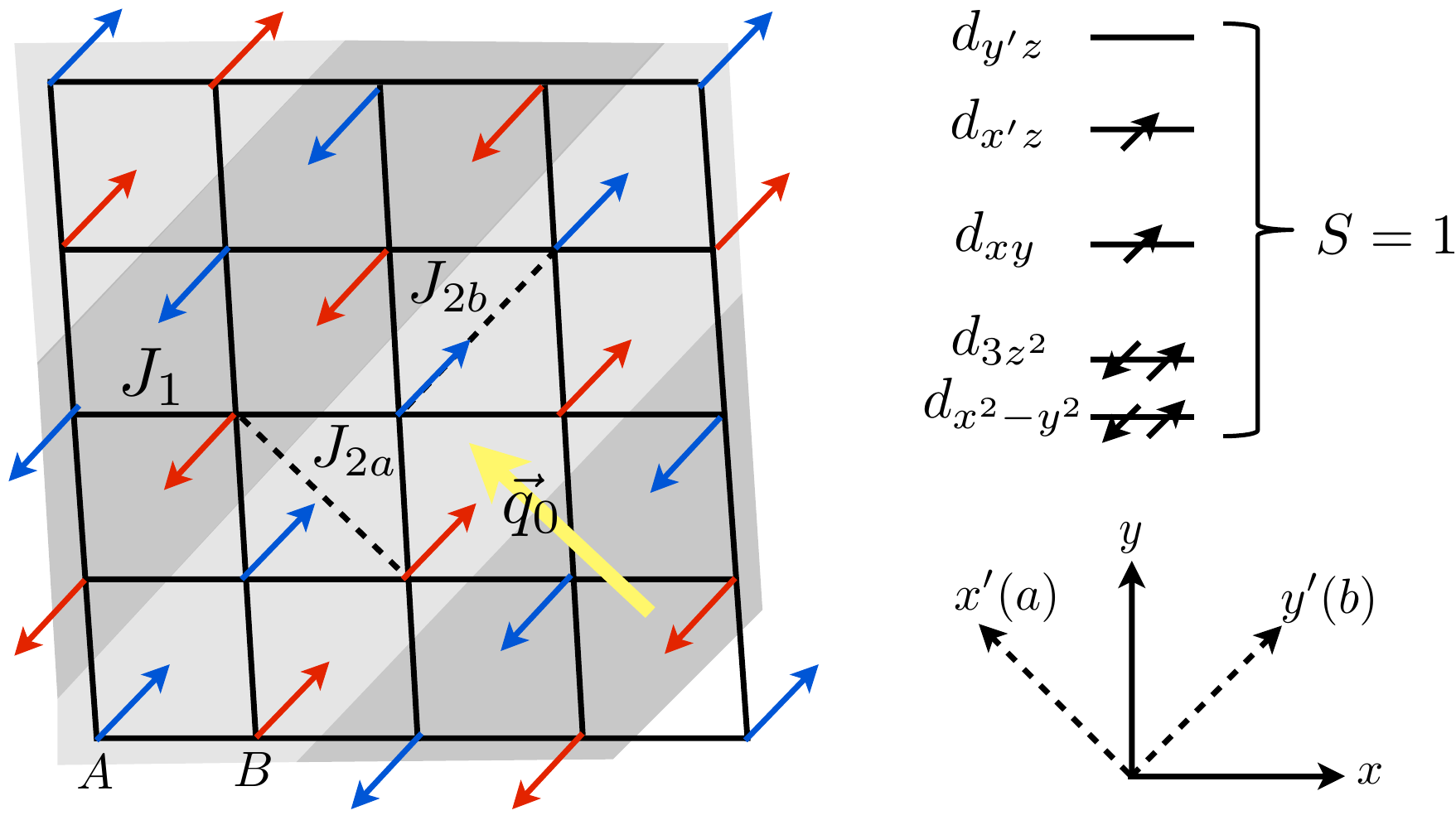}
\caption{ 
Left: bicollinear AFM state with spins locked by the single-ion
  anisotropy to the $b$ crystal axis, and exchange couplings of the
  model Eq.~\eqref{ham} indicated; upper right: the local electron
  configuration of Fe$^{2+}$ in the orthorhombic phase; lower right:
  the choice of coordinates, with the c-axis out of the plane.}
\label{fig2}
\end{figure}

At low temperatures, Fe$_{1+y}$Te distorts from a tetragonal to a weakly 
monoclinic structure at low Fe dopings and to an orthorhombic structure 
at high Fe dopings. Since the monoclinic distortion is fairly weak, with    
$\beta \simeq 89.2$ degrees~\cite{Bao09}, we neglect it for simplicity and 
take the low temperature crystal structure to be orthorhombic for 
all Fe dopings. At low temperature orthorhombic phase, the
crystal is elongated along $a$ crystal axis and compressed along $b$ 
crystal axis, {\it i.e.} $a>b$. Here $a$ and $b$ are lattice constants along
corresponding lattice directions.  As shown in Fig.~\ref{fig2}, 
the orthorhombic distortion lifts the degeneracy of $d_{y'z}$ and 
$d_{x'z}$ orbitals of Fe$^{2+}$ that is present
in the high temperature tetragonal phase (with $a=b$). 
The lower $e_g$ doublets are both doubly occupied and the upper 
$d_{xy}$ and $d_{x'z}$ orbitals are singly occupied, 
forming a local spin moment $S=1$\cite{Turner09}.

%--------------------------------------------------------------
%---------------------------------------------------------------

As discussed in Ref.~\onlinecite{Turner09}, because of the single
occupancy of $d_{xy}$ and $d_{x'z}$ orbitals, the exchange
interaction, $J_{2a}$, along $a$ (or equivalently $x'$) axis is
expected to be antiferromagnetic.  On the other hand, along the $b$
($y'$) axis that is more metallic we take to be ferromagnetic with
exchange, $J_{2b}$, via a double exchange of the extra electron on the
anisotropic upper $d_{y'z}$ orbital due to Fe (self-) doping.  Such
ferromagnetic exchange can also arise from the nearly 90 degree
exchange path~\cite{Fang09}. In addition to next nearest neighbor
(NNN) exchanges, $J_{2a}$ and $J_{2b}$, we include an
antiferromagnetic exchange $J_1$ between nearest neighbors (NN), that
for simplicity we take to be the same along $x$ and $y$ directions.
The orthorhombic single-ion anisotropy is allowed by symmetry and
microscopically arises from the second order contribution of the
spin-orbit coupling. Thus, we take the full Hamiltonian to be
\begin{eqnarray}
{\mathcal H} &=& J_1 \sum_{\langle ij \rangle} {\bf S}_i\cdot{\bf S}_j
+J_{2a}\sum_{\la\la ij \ra\ra_a}  {\bf S}_i\cdot{\bf S}_j
-J_{2b}\sum_{\la\la ij \ra\ra_b}  {\bf S}_i\cdot{\bf S}_j \nonumber \\
& & + \sum_i  \left[ D_a (S_i^a)^2 - D_b (S_i^b)^2 \right],
\label{ham}
\end{eqnarray}
in which, the last term is the single-ion spin anisotropy allowed by
the orthorhombic symmetry, with $D_a, D_b > 0$, so that $b$ is the
easy axis. Because all the interesting spatial variation takes place
in the $ab$-plane of FeTe, above and throughout the paper we focus on 2D
case of a single plane.

We first treat this microscopic model classically, together with the
linear spin-wave analysis appropriate at low temperature inside the
ordered states.  We supplement this with a hydrodynamic theory more
appropriate at high temperatures in Sec.~\ref{sec4}, that allows us to
compute the dynamic structure factors measured via inelastic neutron
scattering in Ref.~\onlinecite{Parshall12}.

\subsection{Mean field analysis}
\label{sec2b}

First we find the ground state of ${\mathcal H}$ for a vanishing spin
anisotropy, $D_a=D_b=0$, treating spins classically. Straightforward
calculation shows that the ground state is a coplanar spin spiral with
\begin{equation}
{\bf S}_i = S [\cos ({\bf k} \cdot {\bf r}_i) \hat{m}_1 +  \sin ({\bf k} \cdot {\bf r}_i )\hat{m}_2],
\label{eq2}
\end{equation}
where $\hat{m}_1$ and $\hat{m}_2$ are two orthogonal unit vectors and
the ordering wavevector ${\bf k} =[k_1,-k_1]=-k_1\hat{\bf a}$ with 
\begin{align}
\cos k_1 = - \frac{J_1}{2J_{2a}}.
\label{cosk1}
\end{align}
Generically this spin spiral is incommensurate for $\frac{J_1}{2J_{2a}} < 1$.  When
$\frac{J_1}{2J_{2a}}\ge 1$, the ground state is the conventional
$[\pi, \pi]$ N\'eel state on the square lattice.  Since in the
experiments\cite{Bao09,Stock11,Rodriguez11,Parshall12} 
$k_1$ is observed to be very close to $\pi/2$,
so we expect $J_1 < J_{2a}$ in FeTe.

The inclusion of the spin anisotropy raises the competition between
the exchange that favors the incommensurate spin-spiral state and the
spin anisotropy that favors collinear states with spins aligned along
the $b$ axis. If $J_1$ is small compared to $J_{2a}$ and $J_{2b}$, the
collinear states should have a ferromagnetic spin configuration along
the $b$ direction and an antiferromagnetic spin configuration along
the $a$ direction.  The resulting collinear state turns out to be the
bicollinear AFM state with the commensurate ordering wavevector $[\pi/2,
-\pi/2]$.  Moreover, for an incommensurate coplanar state, the spin
anisotropy locks the spin spiral onto the easy $bc$
plane. Furthermore, with the presence of the spin anisotropy the $b$
and $c$ directions are no longer equivalent, and the spin spiral can
lower its energy by stretching the spin component along $b$ direction
and shrinking the spin component along $c$ direction. The resulting
spin state no longer has a simple form like Eq.~\eqref{eq2}. When the
hard spin constraint is softened after fluctuations are included, the
spin order may be approximated as an elliptical spin spiral.  Such a
magnetic order is actually observed in the Fe$_{1+y}$Te samples with
$y \gtrsim 0.12$\cite{Rodriguez11}. For the purpose of this section,
we merely approximate the incommensurate ordered state as a circular
spin spiral.
 
\begin{figure}[h]
\includegraphics[width=7cm]{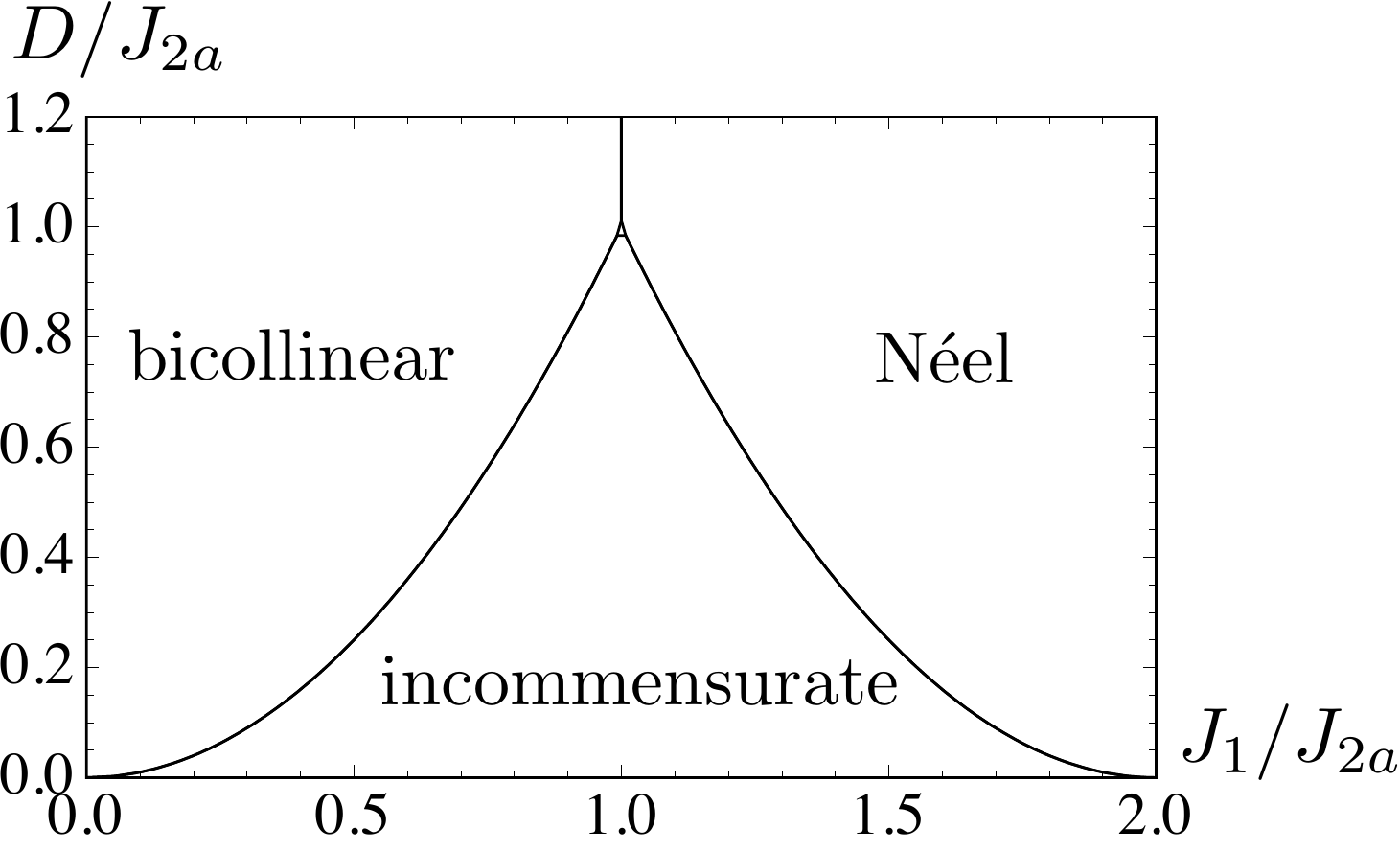}
\caption{
The classical mean field phase diagram with $D_a = D_b \equiv D$ and $J_{2a} > 0$. 
}
\label{fig3}
\end{figure}

Now we consider the three candidate ground states: the bicollinear AFM
state with spins aligned along $b$ direction, the incommensurate state
with a {\it circular} spin spiral in $bc$ plane, and the N\'eel state
with spins aligned along $b$ direction. The classical energies of the
three states are listed in Table~\ref{tab1}. Comparing these energies,
we obtain a phase diagram depicted in Fig.~\ref{fig3}. One should note
that, because we approximate the incommensurate state as a circular
spin spiral, the actual region for the incommensurate state should be
larger than the one in Fig.~\ref{fig3}. Moreover, from
Table~\ref{tab1}, the $J_{2b}$ exchange does not differentiate among
the three states, whether it is small or large.

\begin{table}[t]
\centering
\begin{tabular}{c|c}
ground state  & classical energy per site \\
\hline\hline
commensurate                   & $-(J_{2a} + J_{2b} + D_b) S^2$ \\
\hline
incommensurate                & $- (J_{2a} + J_{2b} + \frac{J_1^2}{ 2 J_{2a} } + \frac{D_b - D_a}{2} ) S^2$ \\
\hline
N\'eel                                & $-(- J_{2a} + J_{2b} + 2 J_1 + D_b) S^2  $ \\
\hline
\end{tabular}
\caption{ The classical energies for three candidate ground states. $S$ is the spin magnitude. }
\label{tab1} 
\end{table} 

\subsection{Linear spin wave theory}
\label{sec2c}

Here we turn our attention to the spin-wave excitations that we study
using the linear spin-wave analysis for the commensurate bicollinear
and the incommensurate coplanar states that are experimentally
relevant. For concreteness, we set $D_a = D_b \equiv D$ in this subsection. 

Since the commensurate bicollinear AFM state is not a proper
state~\cite{Messio11}, we label the two sublattices of the square
lattice as A and B (see Fig.~\ref{fig2}).  The spin orientations are
then parameterized as
\bse
\begin{align}
\label{eq3}
\hat{n}_{\text{A},i} = &\, (-)^{(x_i-y_i)/2} \hat{b},
\\
\hat{n}_{\text{B},i}  = & \, (-)^{(x_i-y_i-1)/2} \hat{b},
\end{align}
\ese
where $x_i-y_i$ is even (odd) for A (B) sublattice. We set the lattice
constants to unity.

The Holstein-Primakoff transformation for the spins is then given by
\bse
\begin{align}
S^{+}_{\mu,i} \equiv &\,  {\bf S}_{\mu,i} \cdot
(\hat{c} + i \hat{n}_{\mu,i} \times \hat{c} )  = \sqrt{2S} a^{\phantom{\dagger}}_{\mu,i}, \\
S^{-}_{\mu,i} \equiv & \,
  {\bf S}_{\mu,i} \cdot (\hat{c} - i \hat{n}_{\mu,i} \times \hat{c} ) = \sqrt{2S} a^{\dagger}_{\mu,i}, \\
&{\bf S}_{\mu,i} \cdot  \hat{n}_{\mu,i} = S - a^{\dagger}_{\mu,i} a^{\phantom{\dagger}}_{\mu,i},
\end{align}
\ese
where $a^{\dagger}_{\mu,i}, a^{\phantom{\dagger}}_{\mu,i}$ are bosonic
creation and annihilation operators with $\mu = \text{A,B}$ the
sublattice index.

For the incommensurate state, it is difficult to parameterize the
coplanar spin order that satisfies the hard spin constraint and
optimizes the energy when the spin anisotropy is present.  However,
experimentally one finds that orthorhombic distortion is reduced and
the incommensurate spin spiral becomes more circular as the Fe doping
level is increased. Thus, to compute the spin-wave dispersion, we
consider the Hamiltonian in the absence of the $b$-axis spin anisotropy and a
circular spiral ground state in $bc$ plane that applies to the regimes of high Fe
dopings.  The spin orientation is then given by
\begin{eqnarray}
\hat{n}_i = \cos (k_1 x_i - k_1 y_i) \hat{b}  + \sin (k_1 x_i - k_1 y_i) \hat{c}.
\end{eqnarray}
This is a proper spin state, and the spin operator can be parameterized as 
\bse
\begin{align}
S^{+}_{i} \equiv & \,
 {\bf S}_{i} \cdot (\hat{a} + i \hat{n}_{i} \times \hat{a} )   = \sqrt{2S} a^{\phantom{\dagger}}_i, \\
S^{-}_{i} \equiv & \,
 {\bf S}_{i} \cdot (\hat{a} - i \hat{n}_{i} \times \hat{a} )   = \sqrt{2S} a^{\dagger}_{i}, \\
& \, {\bf S}_{i} \cdot  \hat{n}_{i} = S - a^{\dagger}_{i} a^{\phantom{\dagger}}_i. 
\end{align}
\ese

\begin{figure}[t]
\includegraphics[width=8cm]{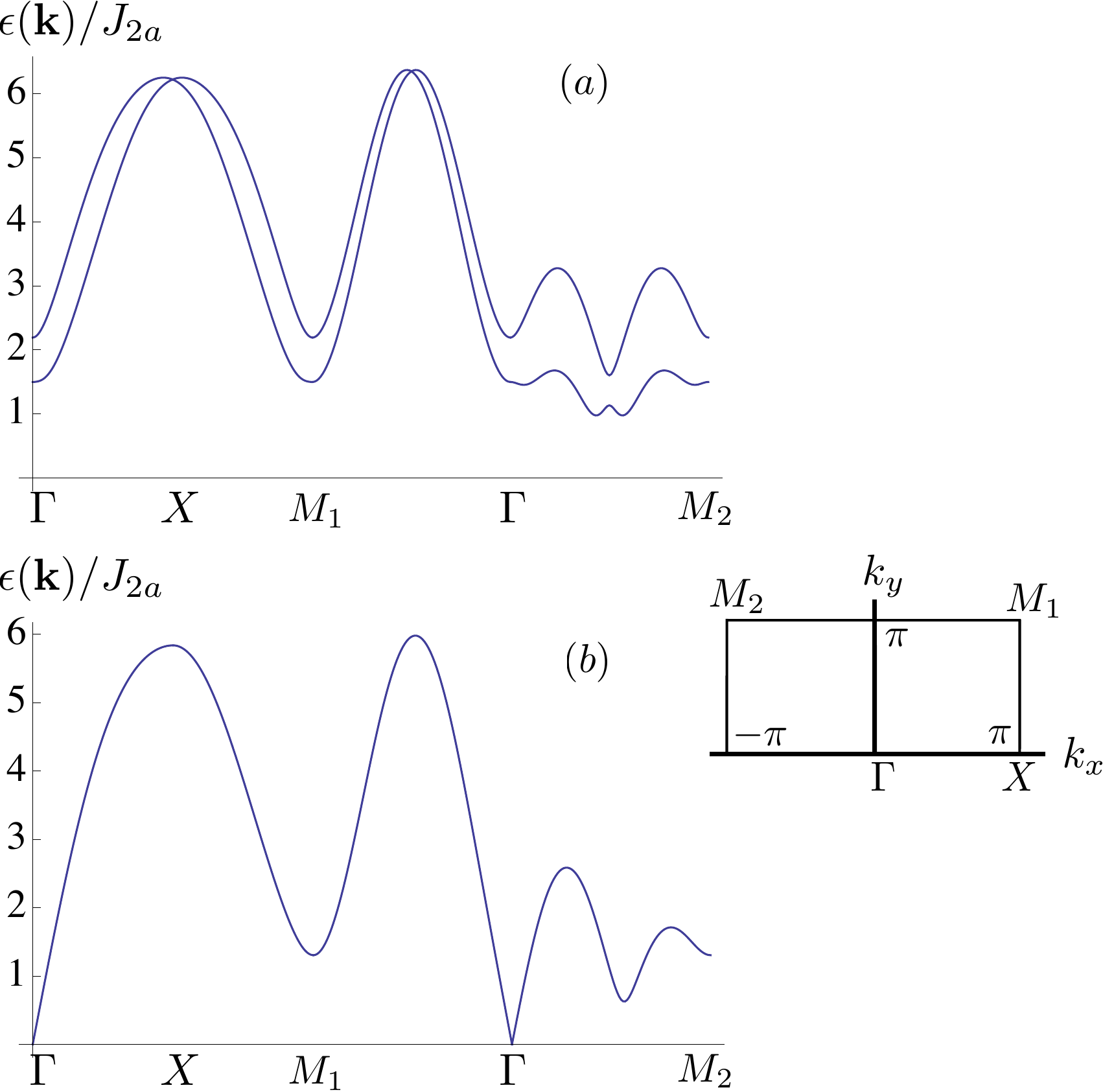}
\caption{The spin wave spectrum for (a) bicollinear state and (b)
  incommensurate coplanar spiral state. In (a), $J_{2a} = J_{2b}, J_1
  = 0.4 J_{2a}, D= 0.2 J_{2a}$, with two dispersions
  corresponding to two spins within the magnetic unit cell.  In (b),
  $J_{2a} = J_{2b}, J_1 = 0.4 J_{2a}, D= 0.05 J_{2a}$.  The
  inset is part of the Brillouin zone with momentum points identified.
}
\label{fig4}
\end{figure}

Plugging these two parameterization into the Hamiltonian Eq.~\eqref{ham}
and keeping to quadratic order in the magnon operators, we obtain the
linear spin-wave Hamiltonians. For the incommensurate spin spiral
state,
\begin{align}
H_{\rm IC}=&
\sum_{\bk}\epsilon_{\bk}^{\rm IC}a_\bk^\dag a^{\phantom\dagger}_\bk
+\nu_{\bk}a_{\bk}^\dag a_{-\bk}^\dag +\nu_{\bk}a^{\phantom\dagger}_{\bk} a_{-\bk}^{\phantom\dagger} \nonumber \\
&+\curN E_{\rm IC}(k_1,-k_1),
\end{align}
with $\curN$ the number of lattice sites, $E_{\text{IC}}$ the ground state
energy of the incommensurate state from Table~\ref{tab1} and
\bse
\begin{align}
\epsilon_\bk^{\rm IC} =&
  J_{2a}  S [ (1 + \cos 2k_1) \cos (k_x-k_y) - 2 \cos 2 k_1  ] \nonumber \\
&+ 2 J_{2b} S [1 - \cos (k_x+ k_y) ] +  J_1 S [ -4 \cos k_1  \nonumber \\
&+ (1 + \cos k_1) ( \cos k_x + \cos k_y )  ] +  D S,
\\
\nu_{\bk} =& 
\frac{1}{2} J_{2a} S ( \cos 2k_1  -1  ) \cos (k_x - k_y) - \frac{1}{2} D S \nonumber \\
& + \frac{1}{2} J_{1} S (\cos k_1 -1) (\cos k_x + \cos k_y) .
\end{align}
\ese
In above expressions $k_1$ is given by Eq.~\eqref{cosk1} which optimizes 
the classical energy. 

For the commensurate bicollinear AFM state, we instead find
\begin{eqnarray}
H_{\rm C} &=& \sum_{\bf k, \mu} \epsilon_{\bf k}^{\rm C} a^{\dagger}_{{\bf k}\mu} 
a^{\phantom\dagger}_{{\bf k}\mu} + \sum_{ {\bf k} } ( \mu_{\bf k} 
a^{\dagger}_{{\bf k}A} a^{\phantom\dagger}_{{\bf k}B} 
+ m_{\bf k} a^{\dagger}_{{\bf k}A} a^{\dagger}_{-{\bf k}B}
\nonumber \\
& +& n_{\bf k} a^{\dagger}_{{\bf k}A} a^{\dagger}_{-{\bf k}A} 
+ n_{\bf k} a^{\dagger}_{{\bf k}B} a^{\dagger}_{-{\bf k}B} 
+ h.c.)
+\curN E_{\rm C}(\frac{\pi}{2},-\frac{\pi}{2})
\nonumber 
\\
\end{eqnarray}
with
\bse
\begin{align}
\epsilon_\bk^{\rm C} =&
2J_{2a} S + 2 J_{2b} S [1-\cos(k_x+k_y) ]  + 3 DS  , \\
\mu_\bk=&m_{\bk}^{\ast}= J_1 S(e^{ik_x}+e^{-ik_y}),\\
n_{\bk}=& J_{2a} S \cos(k_x-k_y) - \frac{1}{2} D S.
\end{align}
\ese
The corresponding spin-wave dispersions are straightforwardly
obtained, and are depicted along the high symmetry lines in
Fig.~\ref{fig4}.  We observe that the spin-wave excitation is gapped
for the bicollinear AFM state as the spin-rotation symmetry is broken
completely by the single-ion spin anisotropy. As expected on general
symmetry grounds, for the incommensurate state we find a gapless
spin-wave excitation spectrum.  Because of the incommensurate nature
of the state, it does not cost any energy to uniformly rotate all the
spins about the $a$-axis and the spectrum remains gapless even if above
analysis is extended to include lattice anisotropy.

\subsection{Effective continuum model}
\label{sec2d}

The microscopic lattice model in Eq.\eqref{ham} gives the ground states
and spin excitation spectrum of Fe$_{1+y}$Te consistent with its
experimental studies~\cite{Rodriguez11, Stock11,
  Lipscombe11, Parshall12}.  To study the low-energy
fluctuations more universally and in more detail, particularly near
continuous phase transitions and beyond mean-field theory, it is
convenient to formulate the system's description using a continuum
Landau-Wilson functional. The latter can be derived from the above
microscopic model using a standard analysis of tracing over the
microscopic spin degrees of freedom in the presence of finite local
magnetization $\langle {\bf S}_i \rangle \sim \boldsymbol{\phi}_i$. 
For a time-independent field configuration (sufficient for our purposes here), 
the harmonic Landau free energy functional is given by,
\begin{equation}
{\mathcal F}[ \boldsymbol{\phi}] 
= \frac{1}{2} \sum_{\bf q} \epsilon_{\bf q}  |\boldsymbol{\phi}_{\bf q}|^2
+ \sum_{{\bf R}_i} \left(\tau_a |\phi_i^a|^2 + \tau_b |\phi_i^b|^2 
+ \tau_c |\phi_i^c|^2 \right),
 \\
\label{eq:freeE}
\end{equation}
where the exchange induced dispersion is well-approximated by
\begin{align}
\epsilon_{\bf q} =& \frac{c_a}{4 q_1^2} (q_a^2 - q_1^2)^2 + c_b q_b^2
\label{vepsilonk}
\end{align}
with a minimum at the incommensurate wavevector, $q_1= 2\pi - 2k_1$ and
$c_{a,b}, \tau_{a,b,c}$ functions of the microscopic parameters
appearing in Hamiltonian \rf{ham} and of temperature, $T$.  Moreover,
the $\tau_{a,b,c}$ terms arise from the easy $ab$-plane (transverse to
$c$-axis) anisotropy together with the single-ion orthorhombic
anisotropy. The experimental orthorhombic lattice phenomenology is
encoded in the $\tau_b < \tau_c < \tau_a$ ordering of these couplings,
with the largest $\tau_a$ confining the spin in the anisotropic
$bc$-plane. The vanishing of $\tau_b$ controls the continuous
transitions from the orthorhombic paramagnet (PM$_{\rm O}$) to the bicollinear
AFM state. In Eq.~\eqref{eq:freeE} we have also changed the basis
from the $xy$ to $ab$ coordinates (see Fig.~\ref{fig2}) using
\begin{eqnarray}
\begin{pmatrix}
q_a \\ q_b
\end{pmatrix}=& R
\begin{pmatrix}
q_x \\ q_y
\end{pmatrix}, \quad
\begin{pmatrix}
x_a \\ x_b
\end{pmatrix}
= \frac{1}{2} R
\begin{pmatrix}
x \\ y
\end{pmatrix}
\end{eqnarray}
with 
\begin{equation}
R = \begin{pmatrix}
1 & -1 \\ 1 & 1
\end{pmatrix}.
\end{equation}
Now the lattice constants in $ab$-coordinates are defined as
a full diagonal length of the lattice in $xy$-coordinates. 
In the following, we will work in the $ab$ coordinates. 

The free energy $\mathcal{F} [\boldsymbol{\phi}]$ captures the
competition between the exchange and the anisotropy. The exchange
dispersion, $\epsilon_{\bf q}$, \rf{vepsilonk} clearly favors an
incommensurate ordering of spins at a wavevector $(q_a,q_b) = (q_1,
0)$, independent of the actual spin orientation, while the ``$\tau$''
anisotropy terms favors a collinear state with magnetization ordered
along $b$.

We note that to capture the spin-lattice commensuration effects above
we were careful to keep the lattice sums in the ``$\tau$'' lattice
anisotropy terms. To go to a complete continuum limit, we coarse-grain
these terms utilizing the Poisson summation formula
\begin{equation}
\frac{1}{v} \sum_{{\bf G}_n} e^{ i{\bf G}_n \cdot {\bf r}} = \sum_{{\bf R}_i} \delta^d ({\bf r} - {\bf R}_i),
\end{equation}
where ${\bf G}_n$ is a reciprocal lattice vector, ${\bf R}_i$ is the
real-space lattice vector with $v$ the volume of the primitive unit
cell.  Applying this to the single-ion anisotropy, we find that the
continuum limit of this free energy is well-approximated by
\begin{eqnarray}
{\mathcal F}_{\text{ani}} [\boldsymbol{\phi}] 
&=& \frac{1}{v} \sum_n \int_{\bf r}\sum_{\mu=a,b,c} 
\tau_{\mu} \phi^{\mu} ({\bf r})^2 e^{i2\pi n x_a} 
\nonumber \\
&\approx&  \frac{1}{v} \int_{\bf r}\sum_{\mu=a,b,c} 
\tau_{\mu} \phi^{\mu} ({\bf r})^2  [1+ 2 \cos (2 q_0 x_a) ],\ \ \
\end{eqnarray}
with $q_0 = \pi$ and the cosine encoding the underlying lattice
discreteness along $a$-axis.  Since we expect magnetic states that are
periodic only along the $a$-axis, in going to the continuum limit we
were safe to neglect the discreteness along other axes.  We further
note that we only kept the lowest reciprocal lattice harmonic, with
higher ones weaker and far from the incommensurate wavevector $q_1$.

\subsection{Magnetoelastic coupling for the single-ion anisotropy}
\label{sec2e}

A semi-microscopic origin of single-ion spin anisotropy discussed
above can be attributed to the structural orthorhombic lattice
distortion that induces spin-anisotropy through spin-orbit
interaction. To this end we consider a general two-dimensional elastic
energy density (here up to quadratic order in the strain) with a
tetragonal symmetry
\begin{align}
\mathcal{H}_{\rm el}^0 = \frac{1}{2}
\left[
K_{11}(u_{xx}^2+u_{yy}^2)+K_{12}u_{xx}u_{yy}+2K_{44} u_{xy}^2
\right]
\label{generalFel}
\end{align}
where $K_{11}$, $K_{12}$, $K_{44}$ are bulk and shear moduli and
$u_{\sigma\sigma'}$ is the elastic strain tensor. Anticipating the
proximity to the tetragonal-to-orthorhombic structural transition,
with principle axes $a$ and $b$ it is convenient to express
$\mathcal{H}_{\rm el}$ in terms of the $ab$ coordinates.  The
transformed strain tensor is then given by
\begin{align}
U^{ab}=
\begin{pmatrix}
u_{aa} & u_{ab} \\ u_{ab} & u_{bb}
\end{pmatrix}
=
R U^{xy} R^{-1}
=
R
\begin{pmatrix}
u_{xx} & u_{xy} \\ u_{xy} & u_{yy}
\end{pmatrix}
R^{-1}
\end{align}
with
\bse
\begin{align}
u_{xx} =& \frac{1}{2}(u_{aa}+u_{bb} + 2u_{ab})\\
u_{yy} =& \frac{1}{2}(u_{aa}+u_{bb} - 2u_{ab})\\
u_{xy} =& \frac{1}{2}(-u_{aa}+u_{bb}).
\end{align}
\ese
Using these relations inside Eq.~\eqref{generalFel} gives
\begin{eqnarray}
{\mathcal H}_{\rm el}^0 &=&
\frac{1}{2}\left[( \frac{1}{2}K_{11}+\frac{1}{4}K_{12}
+\frac{1}{2} K_{44} ) (u_{aa}^2+u_{bb}^2)
 \right.
\nonumber \\ 
&& \left.
+(K_{11}+\frac{1}{2}K_{12}-K_{44})
u_{aa} u_{bb}
+(2K_{11}-K_{12})u_{ab}^2\right] 
\nonumber 
  \\
&\equiv&
\frac{1}{2}\left[K'_{11}(u_{aa}^2+u_{bb}^2)
+K'_{12}u_{aa}u_{bb}
+K'_{44}u_{ab}^2\right]
\end{eqnarray}
with transformed bulk and shear modulus $K'_{11}$, $K'_{12}$, and
$K'_{44}$.  For the tetragonal to orthorhombic transition $u_{xy} \neq
0$ and $u_{xx}=u_{yy}=0$.  Equivalently, in the $ab$ coordinate system
$u_{aa}=-u_{bb}=u_0$ and $u_{ab}=0$, reducing the elastic energy to
\begin{equation}
{\mathcal H}_{\rm el}^0 
= (K'_{11}-\frac{1}{2} K'_{12})u_0^2 
\equiv  K_{44} u_0^2
\end{equation}
In the presence of the orthorhombic distortion, the magneto-elastic
coupling is given by
\begin{eqnarray}
{\mathcal H}_{\rm me} 
&=& \alpha S^{\mu} U_{\mu\nu} S^{\nu}+ g_{12} u_{xy}^2 {\bf S}^2 \nonumber  \\
&=& \alpha u_{aa} (S^a)^2 +\alpha u_{bb} (S^b)^2
+g_{12} u_{xy}^2  {\bf S}^2.
\end{eqnarray} 
Then, including elastic nonlinearities for the orthorhombic strain
$u_0$, the full magneto-elastic Hamiltonian in $ab$ coordinates is
given by,
\begin{equation}
{\mathcal H}_{el} = K_{44} u_0^2+\frac{\lambda}{4}u_0^4
+\alpha u_0 [ (S^a)^2 - (S^b)^2]
+g_{12}u_0^2 {\bf S}^2,
\end{equation}
which, after a structural transition to the orthorhombic state
(characterized by $u_0>0$) leads to the single-ion anisotropy of the
previous section, with $D_a=D_b = \alpha u_0$. Due to the planar
geometry of FeTe, a $c$-axis spin anisotropy is also allowed by
symmetry, both in the tetragonal and orthorhombic phases, and in the
latter state generically leads to $u_{aa} \neq -u_{bb}$ and therefore
$D_a \neq D_b$.

In the above, we demonstrated that an orthorhombic lattice distortion
induces the single-ion spin anisotropy through the magnetoelastic
coupling. More generally, the spin anisotropy, lattice distortion, and
the orbital degrees of freedom should all couple together. In
particular, a Jahn-Teller mechanism is suggested to lift the orbital
degeneracy and induce the lattice distortion. Such a Jahn-Teller
effect can thus also generate the single-ion spin anisotropy.

\section{Magnetostructural transition}
\label{sec3}

We now turn to a mean-field analysis of the magnetostructural
transition of FeTe based on the above Landau theory, similar to Paul
{\it et al}~\cite{Paul11}.

In Fe$_{1+y}$Te, the structural transition from tetragonal to
orthorhombic (for high doping) or monoclinic (for low doping) is
accompanied by the magnetic transition and as a result is naturally
first order.  However, in other iron pnictides the structural
transition is observed to precede the magnetic transition.  As we
demonstrate below, our model captures both possibilities depending on
the value of Landau parameters.

We begin with the Ginzburg-Landau free energy density, $\curF_{\rm GL}$,
\bse
\begin{align}
\curF_{\rm GL}=& \curF_{\rm M} +\curF_{\rm E} + \curF_{\rm ME}, \\
\curF_{\rm M}=& \bphi\cdot\hepsilon_0\cdot\bphi + \tau\ve\bphi \ve^2
+ \frac{g}{2}\ve\bphi\ve^4 , \\
\curF_{\rm E}=& \frac{B}{2}u_{xy}^2 +\frac{\lambda}{4} u_{xy}^4, \\
\curF_{\rm ME}=&\alpha [ (\phi^a)^2-(\phi^b)^2]u_{xy}
+g_{12}\ve\bphi \ve^2\uxy^2 ,
\end{align}
\ese
with all phenomenological couplings positive except $\tau$ and $B$
which can change sign at the structural and magnetic transitions.

As derived in the microscopic analysis of the lattice model the
dispersion exhibits a minimum at a finite momentum $q_1$ along the
$a$-axis and in a continuum is well approximated by
\begin{align}
{\hat\epsilon_0} =& \frac{c_a}{4q_1^2} (-\partial_a^2 -q_1^2)^2 -c_b\partial_b^2-c_c\partial_c^2.
\end{align}
Thus we parameterize the magnetic order by a spiral in the $b$-$c$ plane
\begin{equation}
\boldsymbol{\phi} ( {\bf r} )= \text{Re} [\big(\psi^b({\bf r})\hat{b} -
i\psi^c({\bf r}) \hat{c}\big) e^{i q_1 x_a}   ],
\label{bSspiral}
\end{equation}
where $\psi^\mu({\bf r})$ is a local complex spiral order parameter,
and, anticipating the $b$-$c$ coplanar and $b$ bicollinear AFM
states, we have taken $\phi^a=0$. Because the magneto-elastic coupling
in the $a$-$b$ plane for $u_{xy}>0$ lowers the effective critical
temperature for the $\psi^b$ component (compared to $\psi^c$),
generically we expect the magneto-elastic transition to be
well-characterized by a single magnetic component $\phi\equiv\psi^b$.

With this complex scalar parameterization, within a mean-field
treatment the saddle point equations are given by
\bse
\begin{align}
0 =&\frac{\partial F_{\rm GL}}{\partial \phi^*}
= \tau\phi+g\ve\phi\ve^2\phi-\alpha\phi u_{xy} +g_{12}\phi\uxy^2\\
0 =&\frac{\partial F_{\rm GL}}{\partial u_{xy}}
= B u_{xy} 
- \frac{\alpha}{2}\ve\phi\ve^2
+g_{12}\ve\phi\ve^2\uxy+\lambda u_{xy}^3.
\end{align}
\ese

Firstly we observe that this general magnetoelastic coupling requires
that a nonzero magnetic order always induces a structural distortions
as it is coupled linearly to it.  Thus, a tetragonal phase with
magnetic order is in principle not allowed, though in any particular
system the orthorhombic distortion can be quite small.  Therefore,
this model generically admits the following three phases:
\begin{enumerate}
\item Tetragonal paramagnet, PM$_{\rm T}$ : $\phi =0$, $u_{xy}=0$,
\item Orthorhombic paramagnet, PM$_{\rm O}$ : $\phi=0$ and $u_{xy}\neq 0$,
\item Orthorhombic commensurate or incommensurate AFM states,
  AFM$_{\rm O}$ : $\phi\neq 0$ and $u_{xy}\neq0$
\end{enumerate}

We now map out the corresponding phase diagram.
The PM$_{\rm T}$ state occupies $B>0$, $\tau>0$ part of the phase diagram.

For $B<0$ and $\tau>0$, the system enters PM$_{\rm O}$ state, characterized by
order parameters,
\bse
\begin{align}
\phi =& 0 \\
u_{xy} =&(-\frac{B}{\lambda} )^{\frac{1}{2}}.
\end{align}
\ese
The PM$_{\rm T}$-PM$_{\rm O}$ phase boundary is therefore given by
$B=0$ and $\tau>0$.

On the other hand, for large $B>0$, $\uxy=0$ is a minimum of
$\curF_{\rm E}$, giving $\tau=0$ as the PM$_{\rm T}$-AFM$_O$ phase
boundary at large positive $B$.

To determine the phase boundaries for smaller $B>0$, 
we eliminate (or equivalently integrate out) the strain $u_{xy}$ 
in favor of $\phi$, via
\begin{align}
\uxy \simeq \frac{\alpha}{2B}\ve\phi\ve^2
\end{align}
thereby obtaining an effective Landau free energy density inside
PM$_{\rm T}$
\begin{align}
\curF_{{\rm PM}_{\rm T}}\simeq \frac{1}{2}\tau\ve\phi\ve^2 
+\frac{1}{4}\left(g-\frac{\alpha^2}{2B}\right)\ve\phi\ve^4+\frac{g_6}{6}\ve\phi\ve^6+\cdots
\end{align}
where $g_6 = \frac{3\alpha^2 g_{12}}{4B^2}$, and we neglected
subdominant terms.  For sufficiently large positive $B$ (such that the
$g>\alpha^2/B$) the PM$_{\rm T}$-AFM$_{\rm O}$ transition remains continuous
at $\tau=0$.  However, for $B<B_c(\tau=0)=\alpha^2/(2g)$ such that the
quartic coupling turns negative, the transition is first-order at
$\tau_c(B)$ determined by
\bse
\begin{align}
\curF_{{\rm PM}_{\rm T}}(\phi_0) =&0 \\
\frac{\partial \curF_{{\rm PM}_{\rm T}}}{\partial \phi^*}\Big\ve_{\phi=\phi_0}=&0
\end{align}
\ese

\begin{figure}[tbh]
\includegraphics[width=8cm]{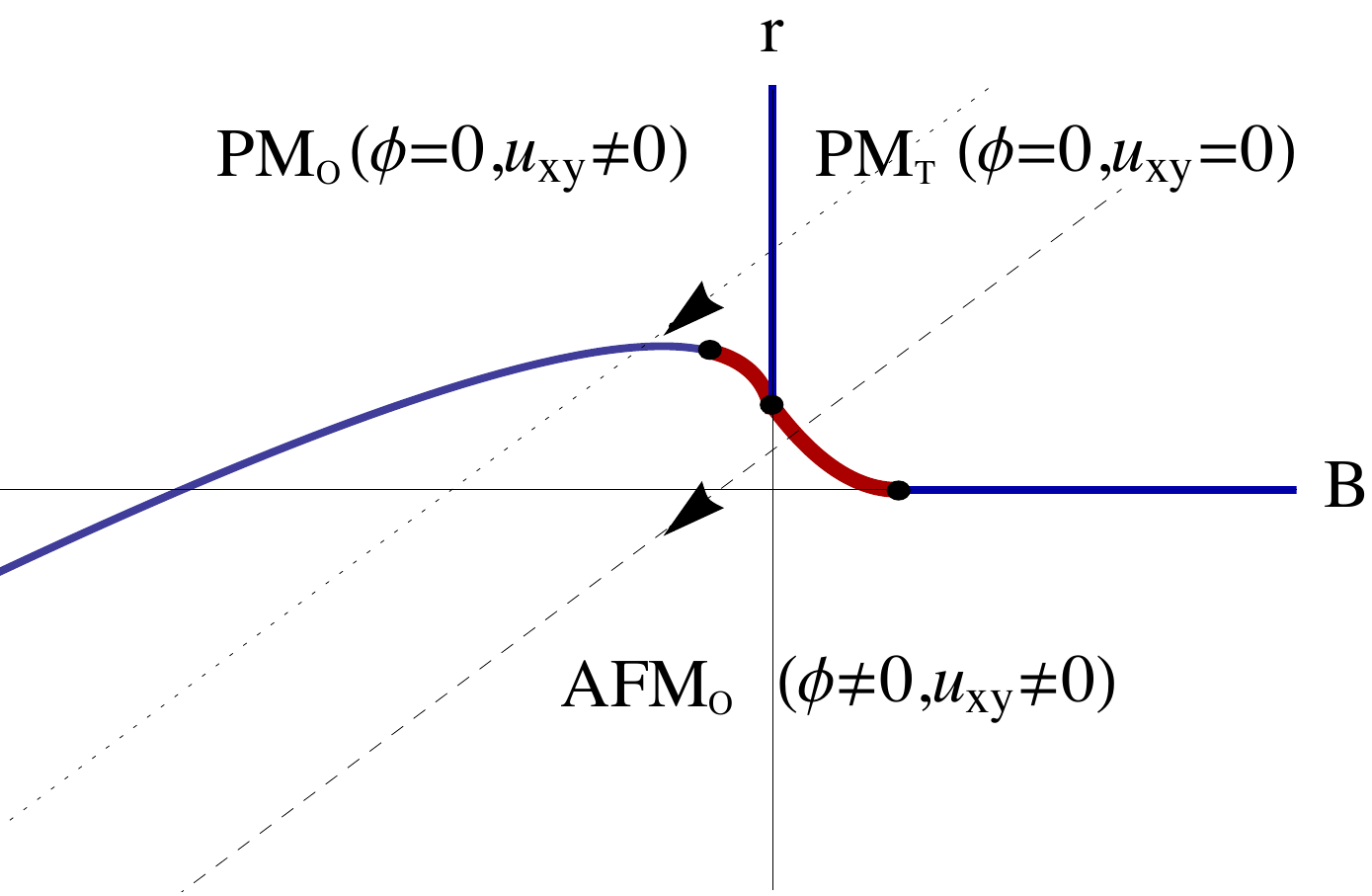}
\caption{ (Color online) The global phase diagram in the reduced
 exchange, $\tau$ and bulk modulus $B$ plane.  For low Fe doping,
  experiments~\cite{Bao09,Li09, Rodriguez11} suggest that
  the reducing temperature follows the dashed curve taking the system
  across the first-order phase boundary (red thick curve) in the
  positive $\tau$-$B$ quadrant, that leads to a simultaneous orthorhombic
  distortion and development of bicollinear AFM order.  The blue
  phase boundaries indicate second-order phase transitions, and in
  FeAs materials are crossed via two distinct, structural and magnetic
  transitions, as indicated by the dotted curve.  }
\label{ms_phasediagram}
\end{figure}

These give
\begin{align}
\ve\phi_0\ve^2 =&\frac{-4\tau_c}{g-\frac{\alpha^2}{B}}
\end{align}
and a first-order transition boundary
\begin{align}
\tau_c(B) =& \frac{1}{16 g_{12}}\left(\alpha -\frac{2gB}{\alpha}\right)^2, \quad \text{for}\quad B>0.
\end{align}
Although this analysis is quantitatively only valid for 
sufficiently large $B>0$, such that elastic nonlinearities remain small,
the qualitative behaviour (upturn in the $\tau_c(B)$ boundary and 
the first-order nature of the transition) persists, as illustrated in Fig.~\ref{ms_phasediagram}.

In contrast, for $B<0$ regime, $u_{xy}$ spontaneously develops a nonzero
expectation value, $u_0$.
For large negative $B$, $u_0$ is determined by balance of
$\uxy^2$ and $\uxy^4$ terms while other terms are small in comparisons.
This gives
\begin{align}
u_0 \simeq \Big(-\frac{B}{\lambda} \Big)^{\frac{1}{2}}
\end{align}
as before and phase boundary is given by (from the saddle point equation),
\begin{align}
\tau_c(B) = \alpha \Big({-\frac{B}{\lambda}} \Big)^{\frac{1}{2}}-\frac{g_{12}B}{\lambda}, \quad \text{for (large) } B<0
\end{align}
However for small negative $B$,
the term linear in $\uxy$ dominates over the $B\uxy^2$ term.
Therefore $u_0 \simeq (\frac{\alpha}{2\lambda}\ve\phi\ve^2)^{1/3}$
and the effective free energy density is given by
\begin{align}
\curF_{{\rm PM}_{\rm O}} \simeq&
\frac{1}{2}\tau\ve\phi\ve^2 + \frac{1}{4}g\ve\phi\ve^4
+\frac{B}{2}\left(\frac{\alpha}{2\lambda}\right)^{\frac{2}{3}}\ve\phi\ve^{\frac{4}{3}}  
\nonumber \\
-& \frac{3\lambda}{4}\left(\frac{\alpha}{2\lambda}\right)^{\frac{4}{3}}\ve\phi\ve^{\frac{8}{3}}.
\end{align}
From this form, we find a first-order transition. Combining the
components analysis, we obtain the phase diagram in
Fig.~\ref{ms_phasediagram}.

For low Fe doping experiments of
Fe$_{1+y}$Te~\cite{Bao09,Li09,Rodriguez11} suggest that
the reducing temperature takes the system across the first-order phase
boundary in the positive $\tau$-$B$ quadrant of phase diagram in
Fig.~\ref{ms_phasediagram}.  This leads to a simultaneous orthorhombic
distortion and development of bicollinear AFM order.  In
contrast, FeAs compounds which exhibit distinct continuous structural
and magnetic transitions, are accommodated by a temperature path
through a continuous phase boundary illustrated in the phase diagram.

\section{Bicollinear-to-spiral low-temperature transition}
\label{sec4}

Focusing on low-temperatures we now examine the nature of transition
between the commensurate bicollinear state and the incommensurate
planar spiral state, that is observed to take place in Fe$_{1+y}$Te at
low temperatures, around the doping level $y\simeq 0.12$.\cite{Rodriguez11}
The transition is driven by a competition between the exchange energy
that prefers a generically incommensurate spiral order at $q_1$ and a
single-ion spin anisotropy that selects a commensurate bicollinear AFM
state at a wavevector $q_0$.  As mentioned earlier, the orthorhombic
distortion is observed to be reduced with increasing Fe doping.  We
therefore expect the single-ion spin anisotropy to also be reduced
with the increased doping level, leading to a commensurate-incommensurate (CI) 
transition when the
spin anisotropy energy drops below the exchange interaction. We
analyze this competition and the resulting CI transition by starting
with the Landau-Wilson free energy functional 
\begin{eqnarray} 
{\mathcal F}[ \boldsymbol{\phi}]
  &\approx&\frac{1}{2} \sum_{\bf q} \epsilon_{\bf q}
  |\boldsymbol{\phi}_{\bf q}|^2 + 
\frac{1}{v} \int_{\bf r} \big[ \tau_{a} |\phi^a|^2
+\tau_{b} |\phi^b|^2+\tau_c |\phi^c|^2 \big]
 \nonumber  \\
  &&\times  [1+2\cos (2 q_0 x_a)],
\label{eq:freeFphiCI}
\end{eqnarray}
for the orthorhombic state, $\tau_b < \tau_c < \tau_a$ derived above.

\subsection{Commensurate-incommensurate transition}
\label{sec4a}

In the bicollinear AFM state, we parameterize the coarse-grained magnetic
order as
\begin{equation}
\boldsymbol{\phi} ({\bf r}) = \psi_0 \cos [q_0 x_a - \theta ({\bf r})] \,\hat{b}, 
\label{mag_pam}
\end{equation}
with the magnetic wavevector $q_0 = \pi$. This parameterization
neglects the subdominant ``massive'' spin fluctuations away from the
easy $b$-axis, focusing on the phase variable $\theta ({\bf r})$. The free
energy density, Eq.~\eqref{eq:freeFphiCI} then reduces to a standard
Pokrovsky-Talapov form\cite{Pokrovsky86}
\begin{equation}
  f [\theta] = \frac{\kappa}{2}  (\partial_{x_a} \theta)^2 
  - \kappa  Q \partial_{x_a} \theta
  - \sigma \cos (2\theta)
\label{eq28}
\end{equation}
where $\kappa = c_a \psi_0^2 /2, \sigma ={-\tau_b \psi_0^2}/{2 v} $
and $Q = q_0 - q_1 $, with $Q \ll q_0$, as is the case in
experiments\cite{Parshall12,Rodriguez11}. In above we dropped 
$\theta$-independent terms and neglected spatial dependence 
transverse to the $a$-axis.

The analysis of this Pokrovsky-Talapov free energy is quite
standard\cite{Pokrovsky86}. We thus omit all technical details, focussing
instead on the qualitative description and results necessary for our
calculation of the structure factor near the CI transition.

Deep in the bicollinear AFM state, the single-ion anisotropy,
$\cos(2\theta)$ locks the phase field $\theta (x_a)$ to zero, thereby
leading to spin order commensurate with the lattice. In this state the
exchange energy is frustrated because the bicollinear AFM order at $q_0$
is not at the minimum $q_1$ of the exchange dispersion. This is
captured by the linear gradient term, $-Q \partial_{x_a} \theta$ that
(when balanced against the exchange $(\partial_{x_a} \theta)^2$) seeks
to induce a constant gradient $Q$ in $\theta$, thereby shifting
magnetic order down to $q_1$.  With increasing iron doping, we expect
the exchange energy to dominate over the single-ion spin anisotropy
and thereby to drive the system through the
commensurate-incommensurate transition at $Q_c$ from the commensurate
bicollinear AFM order at $q_0$ to an incommensurate coplanar spiral order
at $q_1$, with
\begin{equation}
Q_c =  \frac{4}{\pi} \sqrt{\frac{\sigma}{\kappa}}.
\label{eq30}
\end{equation}
The CI transition critical point is defined by the condition that the
energy of a single domain wall in the commensurate state (a spin
``soliton'') vanishes at $Q_c$\cite{Pokrovsky86}.  

At low temperatures, we can ignore thermal fluctuations and
approximate $c_a \approx J_{2a}, Q \approx J_1/J_{2a}$ and $\tau_b \approx -
D_b$, leading to $Q_c \approx (4/\pi) (D_b/J_{2a})^{1/2}$. The system
then develops an incommensurate spin state for
\begin{equation}
D_b \lesssim \frac{\pi^2 J_1^2}{16 J_{2a} }\equiv D_b^{CI}.
\end{equation}

For $Q > Q_c$, the system enters the incommensurate state. At low
temperature, inside the magnetically ordered orthorhombic state this
is captured in terms of a proliferation of $\pi$-solitons in
$\theta(x_a)$, that thereby transitions from its zero value in the
bicollinear AFM state to a periodic array of $\pi$-solitons. Simple
dimensional analysis dictates that a soliton is characterized by a
width $\xi=\sqrt{\sigma/\kappa}$ across which the phase $\theta(x_a)$
advances by $\pi$. For a soliton spacing $d(Q)$, on average this
induces a linear tilt of $\theta$ with $x_a$,
\begin{equation}
  \langle \partial_{x_a}\theta\rangle \equiv \delta q(Q) \approx \pi/d,
\label{thetaAve}
\end{equation}
that corresponds to a shift in the average spin density-wave
wavevector $q(Q) = q_0 - \delta q(Q)$ from $q_0$ down to $q_1$.  The
soliton density $\delta q(Q)$ grows with $Q$ from $0$, asymptotically
approaching $q_0-q_1$ deep in the incommensurate state. At finite
temperature its monotonically increasing functional form exhibits
three regimes of $Q$\cite{Pokrovsky86}.

Because generically a $c$-component of $\boldsymbol{\phi}$ is nonzero
inside the coplanar incommensurate spiral phase, our above
parameterization of $\boldsymbol{\phi} ({\bf r})$, \rf{mag_pam} is
technically incomplete. It thereby describes a CI transition from the
commensurate bicollinear AFM state to incommensurate {\em collinear}
(rather than coplanar) state. At low temperatures we expect the hard
spin constraint to play an important role, and drive the system to
develop a $c$-component of $\boldsymbol{\phi}$. Thus, the transition
is to an elliptical incommensurate planar spiral state.  However,
because close to the CI transition the coplanar incommensurate
regions and the associated phase slips are confined inside solitons
(that are dilute), we expect this approximation of neglecting
$c$-component of $\boldsymbol{\phi}$ to be adequate. Namely, we expect
that near CI the full solution is characterized by a highly eccentric
elliptical polarization with $\phi^c\ll\phi^b$, which is observed in
experiments.\cite{Rodriguez11}

\subsection{Static spin structure factor near the
  commensurate-incommensurate transition}
\label{sec4b}

\begin{figure}[tbh]
\includegraphics[width=8cm]{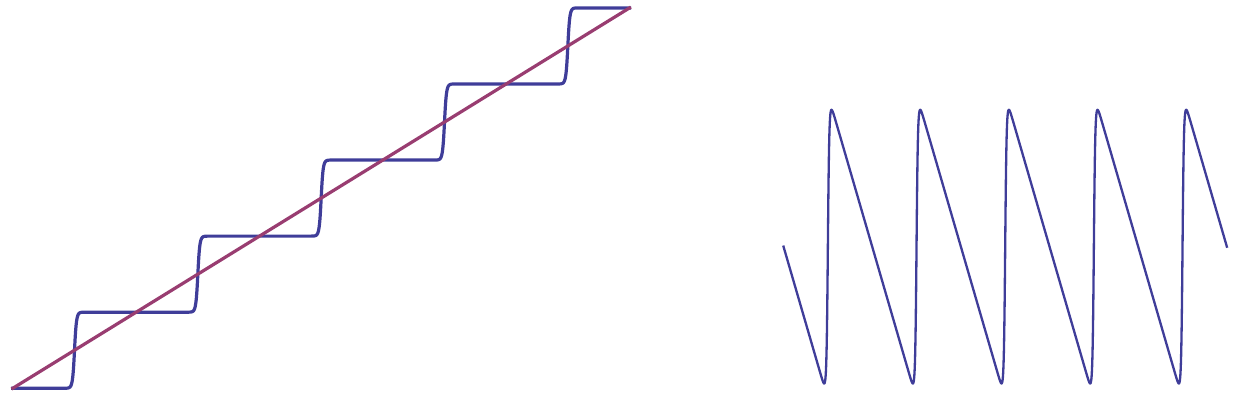}
\caption{ Left: $\theta(x)$ (Elliptic integral of the second kind)
  displaying a train of domain walls in the incommensurate state, just
  above $Q_c$, as well as the average tilted form ${\bar
    \theta}(x)=q x$ (in red).  Right: A train of domain walls in the
  incommensurate state, $\delta\theta=\theta(x)-{\bar \theta}(x)$,
  just above $Q_c$, after ${\bar \theta}(x)=q x$ has been subtracted.
}
\label{fig5}
\end{figure}

We now use above analysis of the commensurate-incommensurate
transition to compute the static spin structure factor,
\begin{equation}
S({\bf q}) \sim \int_{{\bf r},{\bf r}'}   e^{- i {\bf q}\cdot  ( {\bf r}  - {\bf r}' )}  
\langle    \phi^b ({\bf r}) \phi^b ({\bf r}')     \rangle .
\end{equation}  
To compute $S({\bf q})$ we use the parameterization Eq.~\eqref{mag_pam}, with
$\theta(x_a)$ inside the incommensurate state given by a soliton array,
\begin{equation}
\theta(x_a)=\delta q x_a + \delta\theta(x_a),
\end{equation}
where the first term is the average phase from Eq.\rf{thetaAve} and
$\delta \theta (x_a) \equiv \theta (x_a) - \delta q x_a$ is of a
saw-tooth form, that can be well-approximated by a
periodically-extended linear function
\begin{equation}
\delta \theta (x_a) 
= -\frac{\pi}{d} x_a, \quad \text{for} \; -d/2 < x \leq d/2,
\end{equation}
as illustrated in Fig.~\ref{fig5}.

We note that at large $Q\gg Q_c$, $\delta\theta$ vanishes, $\delta
q(Q)\to Q=q_0- q_1$, giving $q_0 x_a - \theta(x_a)\approx q_1 x_a$
(see Eq.\rf{mag_pam}). This reduces the ordering wave vector from
the commensurate $q_0$ to the incommensurate $q_1$ wavevector. Thus,
in this asymptotic limit the spin structure factor displays magnetic
Bragg peaks at integer multiples of $q_1$.

For the intermediate values of $Q > Q_c$ at zero temperature, the
static structure factor is simply given by a Fourier transform 
that includes the soliton contribution,
$e^{i\delta\theta(x_a)}$, that can be easily evaluated in the above
linear (saw-tooth) approximation.  Taking $x_a = d n + x$, we
find
\begin{eqnarray}
&& S( q_a,q_b =0 ) 
\nonumber 
\\
&\propto & | \phi^b(q_a,q_b =0)  |^2 \nonumber
\\
&\simeq &| \sum_{n\in {\mathcal Z}}
e^{-i(q_a-q_0 + \delta q)d n} \int_{-d/2}^{d/2}dx\;e^{-i(q_a-q_0 +
\delta q - \pi/d)x}  |^2
\nonumber 
\\
&\simeq& \sum_{p \in {\mathcal Z}} \frac{1}{ (2p -1)^2}\delta(q_a-q_0 +
\delta q-\frac{2\pi}{d}p)
\end{eqnarray}
Thus, as anticipated, just above $Q_c$ the appearance of solitons
above the CI transition leads to a characteristic scattering
consisting of a sequence of Bragg peaks at $q_p=q_0 - \delta q + 2\pi
p/d$, with the amplitude $A_p = 1/(2p -1)^2$.

At a finite temperature, fluctuations about $\theta_s(x)$ need to be
included.  However, because these are described by an $xy$-model
(since spatial rotational invariance is explicitly broken), we expect
that in 3d these fluctuations are finite and lead to a nonvanishing
Debye-Waller factor suppressing the amplitude of the Bragg peaks but
leaving them sharp at the limit of the resolution (or with a finite
width set by disorder).  Generalization of this analysis to a dynamic
structure function may be of interest in comparing with inelastic
neutron scattering, but will not be performed here.

We next turn our attention to the static and dynamic structure factors
near the finite temperature paramagnetic to magnetic transition, with
the goal to understand the aforementioned neutron scattering
phenomenology of the Fe$_{1+y}$Te observed by Parshall {\it et al}
\cite{Parshall12}.

\section{Static and dynamic spin structure factors near the magnetic
  transition}
\label{sec5}

We now study the spin hydrodynamics of Fe$_{1+y}$Te in the ordered
($T<T_N$) and paramagnetic ($T >T_N$) phases near the
finite-temperature transition at $T_N$ into the magnetic ordered
states. Our motivation here is to understand the anomalous behavior of
the dynamical spin structure factor of Fe$_{1+y}$Te recently observed
by Parshall, {\em et al.}  in the inelastic neutron scattering
experiments near $T_N$\cite{Parshall12}. Studying Fe$_{1.08}$Te,
which exhibits a commensurate bicollinear AFM order and an associated
Bragg peak at lower temperatures, they found that above $T_N=67.5K$,
the inelastic scattering is peaked at an incommensurate wavevector and
zero frequency\cite{Parshall12}. At the magnetic transition, this
incommensurate diffuse peak precipitously shifts to a commensurate
wavevector characteristic of the bicollinear AFM state.  Furthermore, in
this ordered state a spin wave excitation gap is observed, consistent
with other experiments~\cite{Stock11}.

With these experiments performed at a relatively high temperature
regime, {\it i.e.} near $T_N$, we will utilize a classical
hydrodynamic description, extending the planar magnet
hydrodynamics\cite{Halperin69,Hohenberg77} to spiral states with single ion anisotropy
appropriate to Fe$_{1+y}$Te, finding qualitative agreement with
experiments~\cite{Parshall12}.

\subsection{Static spin structure function in orthorhombic
  paramagnetic state }
\label{sec5a}

Before turning to the calculation of the dynamical spin structure
factor, it is instructive to compute the static structure factor in
the orthorhombic paramagnetic state (PM$_{\rm O}$) just above the
transition to the bicollinear planar spiral state. We note that,
although in Fe$_{1+y}$Te this PM$_{\rm O}$ phase has not been observed
(as it undergoes a direct first-order transition from tetragonal
paramagnet (PM$_{\rm T}$) to magnetically ordered orthorhombic phase),
generically PM$_{\rm O}$ is allowed by symmetry and has been observed
in other materials~\cite{Kim11,Prokes11}.  The static spin structure
factor elucidates the competition between the incommensurate spiral
state selected by the exchange interaction and the commensurate state
imposed by the single-ion anisotropy.

\begin{figure}
\includegraphics[width=6.5cm]{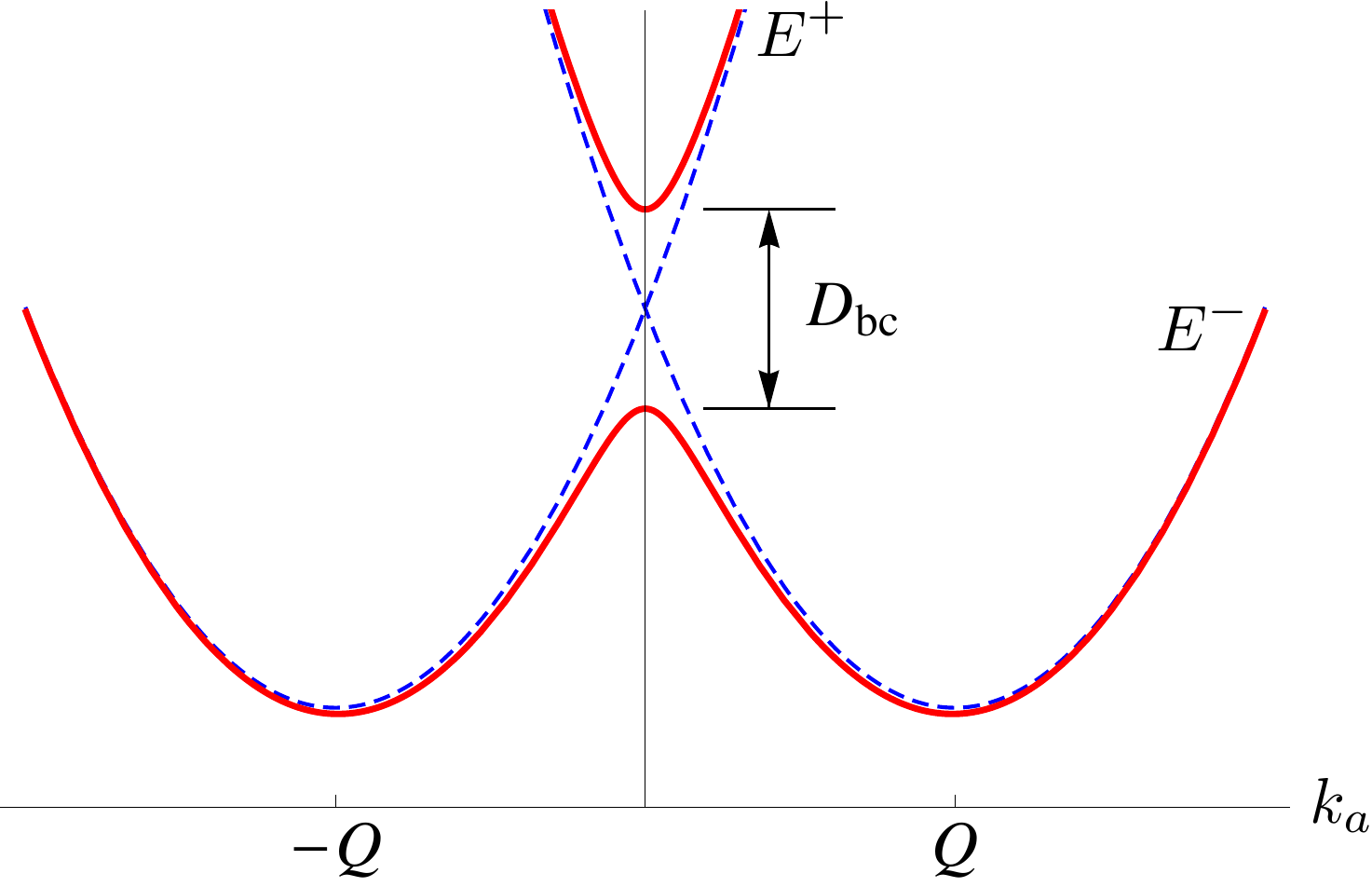}
\caption{The spectra showing two shifted parabolas (dashed curves)
  hybridized (most strongly at their crossings) in proportion to the
  single-ion anisotropy coupling $D_{bc}$.}
\label{gap_static}
\end{figure}

To compute the static spin structure factor, we utilize the free
energy in Eq.~\eqref{eq:freeE}.  Here, we reparameterize the magnetic
order parameter as
\begin{equation}
  \boldsymbol{\phi} ( {\bf r} )= \text{Re} [ \psi({\bf r}) (\hat{b} - i \hat{c} ) e^{i q_1 x_a}   ],
\label{bSspiral}
\end{equation}
where for simplicity we assumed circular polarization taking
$\psi^b({\bf r})= \psi^c({\bf r})\equiv\psi({\bf r})$ as a local
complex spiral order parameter and took $\phi^a=0$. With this
parameterization and in the continuum limit, the free energy in
Eq.~\eqref{eq:freeE} reduces to
\begin{eqnarray}
{\mathcal F} = \frac{1}{2} \sum_{\bf k}{\tilde\epsilon }_{\bf k} |\psi|^2  
- \frac{1}{2} \int_{\bf r} D_{bc} \Big(  \psi^2 e^{2i Q x_a} + c.c.   \Big)
\end{eqnarray}
with ${\bf k} = {\bf q} - q_1\hat{\bf a}$ measured relative to the
incommensurate momentum $q_1\hat{\bf a}$ set by the spin dispersion
minimum, $\tilde\epsilon_{\bf k} = c_a k_a^2 + c_b k_b^2 + \tau,
\tau = \tau_b + \tau_c$, and $D_{bc} = \frac{1}{2v} (\tau_c -
\tau_b)$.  Using $\psi({\bf r}) = \frac{1}{\sqrt{A}} \sum_{\bf k}
\psi_{\bf k} e^{i {\bf k}\cdot {\bf r}}$, standard diagonalization
gives,
\begin{equation}
{\mathcal F}    =  \frac{1}{2} \sum_{\bf k} 
                            \big[   E_{\bf k}^+ |\psi^+_{\bf k} |^2  + E_{\bf k}^- |\psi^-_{\bf k} |^2    \big]
\end{equation}
where
\begin{align}
\begin{pmatrix}
\psi_{{\bf k}+{\bf Q} } \\ \psi_{-{\bf k}+{\bf Q} }^{\ast}
\end{pmatrix}
&=
\begin{pmatrix} u_{\bf k} & -v_{\bf k}^{\ast} \\ v_{\bf k} & u_{\bf k}^{\ast} \end{pmatrix}
\begin{pmatrix} \psi_{{\bf k}}^+ \\ \psi_{{\bf k}}^- \end{pmatrix} 
=
\begin{pmatrix}
u_{\bf k}\psi_{{\bf k}}^+ - v_{\bf k}^{\ast}\psi_{{\bf k}}^- \\
v_{\bf k}\psi_{{\bf k}}^+ + u_{\bf k}^{\ast}\psi_{{\bf k}}^-
\end{pmatrix}
\end{align}
with ${\bf Q} = Q \hat{a}$. The coefficients, $u_{\bf k}$ and $v_{\bf k}$, are given by
\begin{align}
u_{\bf k} = \frac{1}{\sqrt{2}} \left(1+\frac{{\epsilon }_{{\bf k}-}}{E_{\bf k}}\right)^{1/2} , \quad 
v_{\bf k} = \frac{1}{\sqrt{2}}\left(1-\frac{{ \epsilon }_{{\bf k}-}}{E_{\bf k}}\right)^{1/2}
\end{align}
with  
\bse
\begin{align}
\epsilon_{{\bf k}\pm} =&  \frac{1}{4}({\tilde\epsilon }_{{\bf k}+{\bf Q} }\pm
{\tilde\epsilon }_{-{\bf k}+{\bf Q}}),\\
E_{\bf k} =&\sqrt{\epsilon_{{\bf k}-}^2+D_{bc}^2},\\
E^\pm_{\bf k} =&\epsilon_{{\bf k}+}\pm E_{\bf k}.
\end{align}
\ese
The hybridization of the spectra via single-ion anisotropy is depicted
in Fig.~\ref{gap_static}.

\begin{figure}
\includegraphics[width=6.5cm]{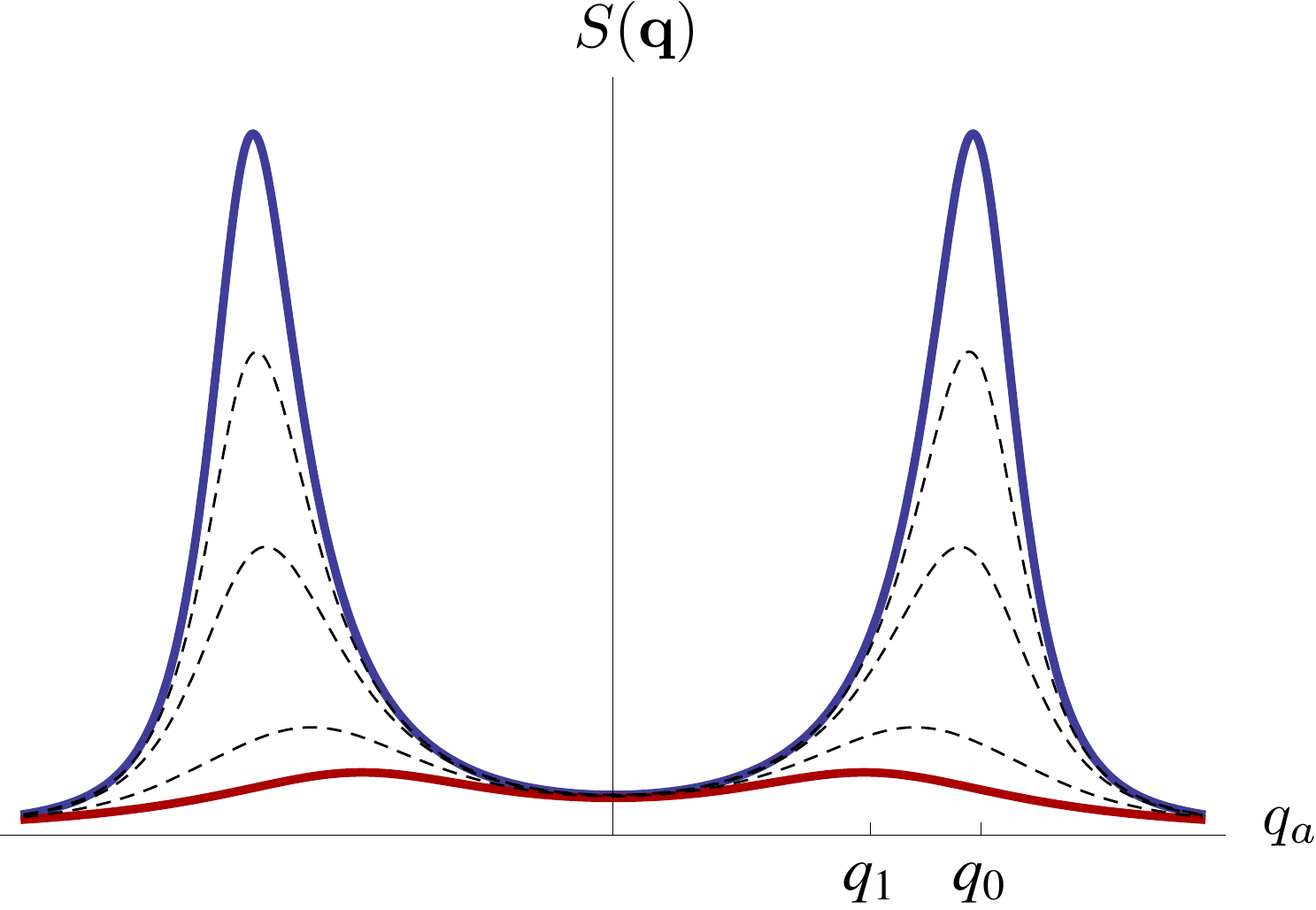}
\caption{ The static structure factor $S({\bf q})$ in the PM$_{\rm O}$
  state, where the shift of peaks from incommensurate wavevector $q_1$
  (bottom curve) to commensurate wavevector $q_0$ (top curve) are
  shown.  The dashed curves are for temperature intermediate values
  with the temperature increases from the top curve to bottom curve.
  The plot is displayed in arbitrary unit.}
\label{shift_static}
\end{figure}

We express the magnetic order parameter $\phi_{\bf q}^b$ in terms of
these normal modes,
\begin{align}
\phi_{\bf q}^b =& 
\frac{1}{2}(v_{{\bf q}+ {\bf q}_0}^{\ast}\psi_{{\bf q}
+{\bf q}_0}^{+\ast} +u_{{\bf q}+{\bf q}_0}\psi_{{\bf q}
+{\bf q}_0}^{-\ast}
\nonumber \\
&+u_{{\bf q}-{\bf q}_0}^{\ast} \psi_{{\bf q}-{\bf q}_0}^{+\ast} 
  -v_{{\bf q}-{\bf q}_0}\psi_{{\bf q}-{\bf q}_0}^{-\ast})
\end{align}
with ${\bf q}_0 = q_0 \hat{a}$. 
Using equipartition for the correlation function of the normal modes
in the paramagnetic state, we obtain the static structure factor
for $T>T_N$,
\begin{eqnarray}
S({\bf q}) &\equiv& \langle \phi_{\bf q}^{b\ast}  \phi_{\bf q}^b \rangle 
\nonumber\\
 &=&\frac{k_BT}{2}\left[
\frac{\ve v_{{\bf q}+\ibq_0}\ve^2}{E_{{\bf q}+\ibq_0}^+}
+\frac{\ve u_{{\bf q}+\ibq_0}\ve^2}{E_{{\bf q}+\ibq_0}^-}
+\frac{\ve u_{{\bf q}-\ibq_0}\ve^2}{E_{{\bf q}-\ibq_0}^+}
+\frac{\ve v_{{\bf q}-\ibq_0}\ve^2}{E_{{\bf q}-\ibq_0}^-} \right]\nonumber\\
 &=&\frac{k_B T}{8}\left[
\frac{\tilde{\epsilon}_{{\bf q}+{\bf q}_0+{\bf Q}}}{E_{{\bf q}+{\bf q}_0}^+E_{{\bf q}+{\bf q}_0}^-}
+\frac{\tilde{\epsilon}_{-{\bf q} + {\bf q}_0+{\bf Q}}}{E_{{\bf q}-{\bf q}_0}^+E_{{\bf q}-{\bf q}_0}^-}
\right].
\label{EQ_static_structure_function}
\end{eqnarray}
The spin structure factor is displayed for varying reduced temperature $\tau$
in the PM$_{\rm O}$ state in Fig.~\ref{shift_static}.

\subsection{Dynamic structure factor in paramagnetic state, PM$_{\rm O}$}
\label{SkwPM}

We now turn to the computation of the dynamic structure factor, first
focusing on the paramagnetic state.  Because the primary
experiments~\cite{Parshall12} of our interest take place at a
relatively high temperature near the thermal transition to the
magnetic state, we utilize a classical hydrodynamic
description~\cite{Halperin69}.  Standard symmetry arguments, together
with nontrivial spin commutation relations that encode precession lead
to a model E hydrodynamics, described by coupled Langevin equations,
\bse
\begin{align}
\partial_t \phi 
=& 
-2\gamma\frac{\delta \mathcal{F} }{\delta \phi^{\ast}}
-i\Gamma \phi \frac{\delta \mathcal{F}}{\delta m} + \zeta 
\label{eomHydrodynamicsPMa}\\
\partial_t m 
=&
 \lambda \nabla^2  \frac{\delta {\mathcal F}}{\delta m}
+2 \Gamma \text{Im} \left( \phi^{\ast} \frac{\delta {\mathcal F}}{\delta \phi^{\ast}}
\right) 
+ \zeta_m
\label{eomHydrodynamicsPMb}
\end{align}
\ese
where $\mathcal{F}$ is from Eq.\rf{eq:freeFphiCI}, $\phi = \phi^b +
i\phi^c$, $m=\phi^a$ are local transverse and longitudinal
magnetization components, and $\Gamma, \gamma, \lambda$ are
hydrodynamic coefficients characterizing the system.  $\zeta=\zeta_b +
i \zeta_c$, $\zeta_m$ are components of the thermal Gaussian noise
characterized by a vanishing mean, $\langle \zeta_\sigma \rangle
=\langle \zeta_m \rangle = 0$ and variances imposed by the
fluctuation-dissipation relation~\cite{Halperin69},
\bse
\begin{align}
  \langle \zeta_\sigma({\bf r},t)\zeta_{\sigma'}({\bf r}',t') \rangle
  =& 2\gamma k_BT \delta_{\sigma\sigma'}\delta({\bf r}-{\bf r}')
  \delta(t-t') 
\label{gaussiannoise_a}\\
  \langle \zeta_m({\bf r},t)\zeta_{m}({\bf r}', t')\rangle =& 
  -2 \lambda k_BT\nabla^2 \delta({\bf r}-{\bf r}')\delta(t-t'),
\label{gaussiannoise_b}
\end{align}
\ese
with $\sigma = b, c$.  The equations consist of dissipative
(relaxational) terms as well as the reactive parts that capture the
spin precessional dynamics, as studied extensively for numerous other
magnetic systems.  The new ingredient here is the spiral nature of the
ordered state and the single-ion pinning anisotropy special to
Fe$_{1+y}$Te.

\begin{figure}
\includegraphics[width=6.5cm]{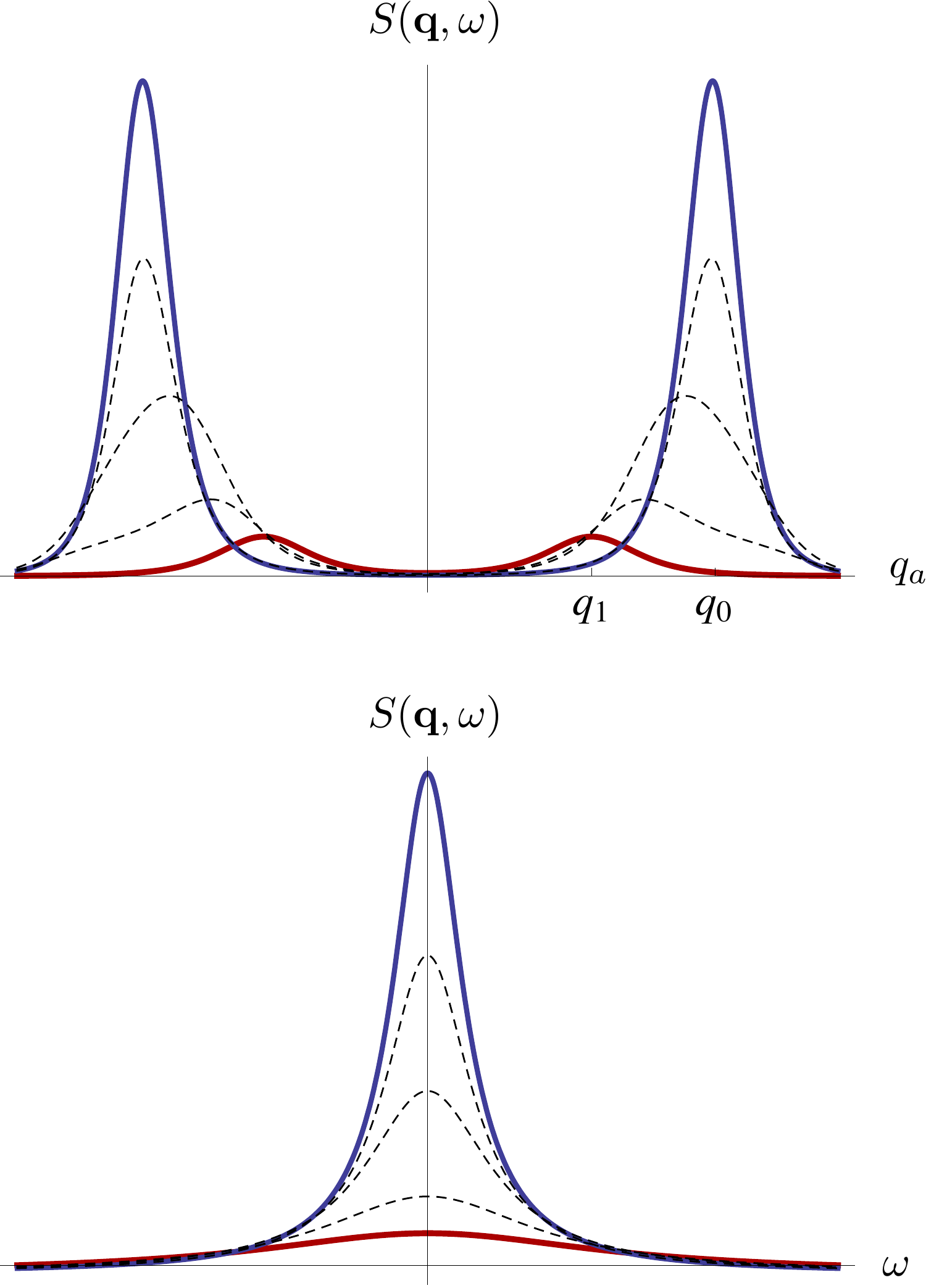}
\caption{The dynamic structure factors $S(\bf q_a,\omega)$ from
  Eq.\rf{EQ_dynamic_structure_function} as a function of $q_a$ (top)
  and $\omega$ (bottom).  As the transition to the SDW at $T_N$ is
  approached it exhibits a shifting and growth of the finite momentum
  peak from the incommensurate state at $q_1$ (red) at high
  temperature toward the commensurate value $q_0$ (top).  The dashed
  lines are the value for intermediate temperatures.  The Lorentzian
  peak at $\omega=0$ represents the relaxational dynamics
  characteristic of the paramagnetic phase.}
\label{shift_dynamic_PM}
\end{figure}

In the disordered paramagnetic state, it is sufficient to work within
the harmonic approximation, where the nonlinear precessional terms can
be neglected.  Thus, the equations of motion considerably simplify to
the model B form\cite{Halperin69}.  The equation for the longitudinal
magnetization $m$ decouples giving a simple diffusive mode for the
spin $a$ component. We focus on the transverse modes encoded in the
complex order parameter $\phi$, that satisfies
\begin{equation}
\partial_t\phi = -\gamma\left[(\hat\epsilon_0 +\tau)\phi -
  4D_{bc}\phi^{\ast}\cos(2q_0 x_a)\right] + \zeta.
\end{equation}
Using Eq.~\eqref{bSspiral}, $\phi = \psi e^{i q_1 x_a}$ for the spin
spiral state parameterization we obtain
\begin{eqnarray}
\frac{\partial}{\partial t}
\begin{pmatrix}
\psi_{{\bf k}+{\bf Q}} \\ \psi^{\ast}_{-{\bf k}+{\bf Q}} 
\end{pmatrix} 
&=&
-\gamma
\begin{pmatrix}
\tilde{\epsilon}_{{\bf k}+{\bf Q}} & -2D_{bc} \\ 
-2D_{bc} & \tilde{\epsilon}_{-{\bf k}+{\bf Q}}
\end{pmatrix}
\begin{pmatrix}
\psi_{{\bf k}+{\bf Q}} \\ \psi^{\ast}_{-{\bf k}+{\bf Q}} 
\end{pmatrix} \nonumber \\ 
&& 
+
\begin{pmatrix}
\zeta_{{\bf k}+{\bf q}_0} \\ \zeta^{\ast}_{-{\bf k}+{\bf q}_0} 
\end{pmatrix}
\label{eqmat1}
\end{eqnarray}
where $\tilde{\epsilon}_{\bf k}= c_ak_a^2+c_bk_b^2+\tau$ and $\zeta_{\bf q}$
is a spatial Fourier transform of $\zeta({\bf r})$ with
\begin{equation}
\langle \zeta_{\bf k}^\ast(t)\zeta_{{\bf k}}(t')\rangle = 
4\gamma k_BT\delta(t-t').
\end{equation}
After Fourier transformation, we find
\begin{equation}
\psi({\bf k}+{\bf Q},\omega) =
C_{\bf k}\Big[(i\omega + \gamma\tilde{\epsilon}_{-{\bf k}+{\bf Q}})
\zeta_{{\bf k}+{\bf q}_0}+2\gamma D_{bc}\zeta_{-{\bf k}+{\bf q}_0}\Big],
\end{equation}
where 
\begin{equation}
C_{\bf k} = [-\omega^2+4\gamma^2(\epsilon_{{\bf k}+}^2
-E_k^2+i\omega \epsilon_{{\bf k}+}/\gamma)]^{-1}.
\end{equation}
From we obtain the dynamic spin structure factor
\begin{widetext}

\begin{eqnarray}
S({\bf q},\omega)&\equiv &\la \phi^b({\bf q},\omega)\phi^b(-{\bf q},-\omega)\ra
\nonumber  
\\ 
&=&
\frac{1}{4}\la \psi({\bf q}-{\bf q}_1,\omega)
\psi^{\ast}({\bf q}-{\bf q}_1,\omega)\ra 
+\frac{1}{4}\la \psi^{\ast}(-{\bf q}-{\bf q}_1,-\omega)
\psi(-{\bf q}-{\bf q}_1,-\omega)\ra
\nonumber
\\
&=&
\frac{\gamma k_B T [ \omega^2+\gamma^2
(\tilde\epsilon_{{\bf q} + {\bf q}_0+\ibQ}^2 + 4D_{bc}^2)]}
{[-\omega^2+4\gamma^2(\epsilon_{{\bf q}+ {\bf q}_0,+}^2-E_{{\bf q}+{\bf q}_0}^2)]^2
+(4\gamma\omega\epsilon_{{\bf q}+{\bf q}_0,+})^2}
+\frac{\gamma k_B T [\omega^2+\gamma^2
(\tilde\epsilon_{-{\bf q}+{\bf q}_0+\ibQ}^2+ 4D_{bc}^2) ] }
{[-\omega^2 + 4\gamma^2(\epsilon_{-{\bf q}+{\bf q}_0,+}^2-E_{-{\bf q}+{\bf q}_0}^2)]^2
+(4\gamma\omega\epsilon_{-{\bf q}+{\bf q}_0,+})^2}.
\nonumber
\\
\label{EQ_dynamic_structure_function}
\end{eqnarray}

\end{widetext}
For a range of parameters, $S({\bf q},\omega)$ is illustrated in
Fig.~\ref{shift_dynamic_PM}. It displays an $\omega=0$ Lorentzian peak
and a shift of its finite ${\bf q}$ peaks from the incommensurate to
commensurate values as observed in experiments by Parshall {\it et
  al}~\cite{Parshall12}.  Consistent with the fluctuation-dissipation
theorem, integration of $S({\bf q},\omega)$ over $\omega$ also gives
the static structure factor obtained directly in Sec.~\ref{sec4a},
Eq.~\eqref{EQ_static_structure_function} and illustrated in
Fig.~\ref{shift_static}.

\subsection{Dynamic structure function in planar spiral state}

\subsubsection{Incommensurate phase}
 
Next we turn to the computation of the dynamic structure factor
inside the {\em incommensurate} planar spiral state. Starting
with the free energy, Eq.~\eqref{eq:freeFphiCI}, and neglecting the
$bc$ plane spin-anisotropy (as it is averaged out in the incommensurate
state), the free energy reduces to
\begin{align}
  \mathcal{F} = \frac{1}{2}\int dx_a
  dx_b \big[\sum_{\mu=b,c}\phi^{\mu} \hepsilon_0 \phi^{\mu}
    +\chi_m^{-1}m^2 \big],
\label{Fspiral}
\end{align}
with $\chi_m$ the $a$-axis magnetic susceptibility. To treat the
dynamics we need to include the $a$-component of the magnetization,
$m$, because it is a field conjugate to the hydrodynamic mode
$\phi=\phi^b + i\phi^c$ and therefore appears in the spin precessional
term in the equations of motion. In the spiral state it appears even
at a harmonic level. This contrasts with the statics
(Sec.~\ref{sec4a}, where $m$ is ``gapped'' and therefore at low
energies can be neglected) and with the disordered state dynamics
(Sec.~\ref{SkwPM}, where in the harmonic approximation $m$ decouples
from the transverse Goldstone modes, $\phi^{b,c}$). The hydrodynamic
equations for the complex $\phi$ order parameter and its conjugate
field $m$ become
\bse
\begin{align}
\partial_t \phi 
=& 
-\gamma\hat\epsilon_0\phi - i \Gamma\chi^{-1}_m m \phi + \zeta,
\label{eomHydrodynamicsSpirala}\\
\partial_t m 
=&  D_m \nabla^2 m 
+ \Gamma \text{Im} \left(\phi^{\ast}\hat\epsilon_0\phi\right) 
+ \zeta_m,
\label{eomHydrodynamicsSpiralb}
\end{align}
\ese
with $D_m = \lambda / \chi_m$. 
Deep within the spiral state, where the magnetization is parameterized
by $\phi = |\psi_0|e^{i\vphi + i q_1 x_a}$ and $\psi_0$ can be taken
to be a constant, the hydrodynamic equations reduce further to linear
coupled equations for the Goldstone mode $\vphi({\bf r})$ and its
conjugate $a$-component of the magnetization, $m({\bf r})$
\bse
\begin{align}
\partial_t\vphi
=& 
- \gamma \tilde{\epsilon}_0 \vphi -{\Gamma} m /\chi_m+ \zeta_\vphi
\\
\partial_t m 
=&
D_m \nabla^2 m + \Gamma|\psi_0|^2\tilde{\epsilon}_0 \vphi+\zeta_m
\end{align}
\ese
with $\tilde{\epsilon}_0 = - c_a \partial_a^2 - c_b \partial_b^2$, and
$\zeta_\vphi = \text{Re}[-i\zeta e^{-i q_1 x_a}]/|\psi_0|$.  These
strongly resemble model-E hydrodynamics, appropriate for our planar
magnetic system~\cite{Halperin69}, with the noise $\zeta_\vphi$
governed by a zero-mean Gaussian statistics, characterized by
\begin{equation}
\\
\la \zeta_\vphi(\ibr,t) \zeta_\vphi(\ibr',t)\ra = 2 \gamma k_B T
\delta(\ibr-\ibr')\delta(t-t')/|\psi_0|^2
\label{eq60}
\end{equation}

After Fourier transform, we obtain,
\begin{eqnarray}
\begin{pmatrix}
\vphi_{{\bf k},\omega} \\ m_{{\bf k},\omega}
\end{pmatrix} 
&\simeq&
{(-\omega^2+\Omega_{\bf k}^2-i\omega D_{\bf k} { k}^2)^{-1}} \nonumber \\
&&
\begin{pmatrix}
-i\omega+ D_m {k}^2 & -{\Gamma}/{\chi_m}\\
\Gamma|\psi_0|^2 \tilde{\epsilon}_{0,{\bf k}} &-i\omega 
+ \gamma \tilde{\epsilon}_{0,{\bf k}}
\end{pmatrix}
\begin{pmatrix}
\zeta^\vphi_{{\bf k},\omega} \\ \zeta^m_{{\bf k},\omega}
\end{pmatrix}
\nonumber \\
\end{eqnarray}
with $\tilde{\epsilon}_{0,{\bf k}} = c_a k_a^2 + c_b k_b^2 $, and
\bse
\begin{align}
\Omega_{\bf k} =& 
\left({\Gamma^2|\psi_0|^2}/{\chi_m}
+\gamma D_m {k}^2\right)^{1/2}\tilde{\epsilon}_{0,{\bf k}}^{1/2},\\
D_{\bf k} =& \gamma \tilde{\epsilon}_{0,{\bf k}} /{ k}^2
+ D_m.
\end{align}
\ese
The poles of the above transfer function (zeros of the characteristic
equation) give the spectrum in the incommensurate spiral state
\bse
\begin{align}
\omega_{\bf k}=&
-\frac{i}{2}(\gamma \tilde{\epsilon}_{0,{\bf k}} +D_m {k}^2) 
\pm\frac{1}{2} \big[
-(\gamma \tilde{\epsilon}_{0,{\bf k}} +D_m { k}^2)^2
\nonumber
 \\
&+4D_m \gamma
\tilde{\epsilon}_{0,{\bf k}} {k}^2
+4{\Gamma^2|\psi_0|^2} \tilde{\epsilon}_{0,{\bf k}} / \chi_m \big]^{1/2}  \\
\simeq &
\pm \Gamma\sqrt{ {|\psi_0|^2\tilde{\epsilon}_{0,{\bf k}}}/{\chi_m}}
-\frac{1}{2}i(\gamma\tilde{\epsilon}_{0,{\bf k}} +D_m { k}^2) \\
\simeq &
\pm \Omega_{\bf k}-\frac{i}{2}D_{\bf k} { k}^2 \\
\simeq &
\pm v_{\hat{\bf k}} k-\frac{i}{2}D_{\bf k} { k}^2
\end{align}
\ese
where $\Omega_{\bf k}
\approx  \Gamma |\psi_0| \sqrt{\tilde{\epsilon}_{0,{\bf k}} /\chi_m } \equiv
v_{\hat{\bf k}} { k}$.

Averaging over noise of Eq.~\eqref{eq60},
we obtain correlation functions which are defined in the following form, 
\begin{equation}
\la \vphi_{{\bf k},\omega}\vphi_{{\bf k}',\omega'}\ra =
C_{\vphi\vphi}( {\bf k},\omega)(2\pi)^3
\delta({\bf k}+{\bf k}')\delta(\omega+\omega'),
\end{equation}
and 
\bse
\begin{align}
C_{\vphi\vphi}
\simeq&
\frac{2k_B T(\gamma\omega^2/|\psi_0|^2
+ \lambda\Gamma^2 { k}^2/\chi_m^2 ) }
{(\omega^2-\Omega_{\bf k}^2)^2+\omega^2 (D_{\bf k} { k}^2)^2 } 
\\
C_{m m} 
\simeq&
\frac{
2k_BT (\gamma \Gamma^2|\psi_0|^2\tilde\epsilon_{0,{\bf k}}^2 
+ \lambda \omega^2 { k}^2)}
{(\omega^2-\Omega_{\bf k}^2)^2+\omega^2 (D_{\bf k} { k}^2)^2}
\\
C_{\vphi m} 
\simeq&
\frac{-i2k_B T \Gamma \omega D_{\bf k} { k}^2}{(\omega^2
-\Omega_{\bf k}^2)^2+\omega^2  (D_{\bf k} { k}^2)^2},
\end{align}
\ese
where above we neglected terms that are subdominant in the
long-wavelength, low-frequency limit.  We note that these correlation
functions are quite similar to those of the planar uniform magnet
obtained in Ref.~\onlinecite{Halperin69}.  This is despite the fact
that here the ordered state is a {\em periodic} planar spiral.
Nevertheless, because in this state the conjugate field $m$ (the
$a$-component of the magnetization) remains uniform, vanishing on
average and locally conserved (as in uniform magnetic states), its
hydrodynamics is qualitatively identical to that of an easy-plane
ferromagnet and a superfluid (model E).

\subsubsection{Commensurate phase}

We now apply above analysis to the commensurate state, where the
pinning anisotropy arising from the orthorhombic distortion plays an
important role.  Its main consequence is pinning of the spiral at a
wavevector $q_0$ commensurate with the lattice, orienting spins along
the $b$-axis and thereby opening the gap in the spin wave spectrum.

By definition, deep in the commensurate phase the pinning is strong,
with the equation of motion given by
\bse
\begin{align}
\partial_t\vphi
=& 
-\gamma\tilde\epsilon_0\vphi
-{\Gamma}m/{\chi_m}
-2\gamma D_{bc}\sin(2\vphi - 2Q x_a)+\zeta_\vphi \\
\partial_t m 
=&
D_m \nabla^2 m +
\Gamma|\psi_0|^2\tilde\epsilon_0\vphi\nonumber\\ 
&+2\Gamma D_{bc}|\psi_0|^2\sin(2\vphi - 2Q x_a) + \zeta_m.
\end{align}
\ese
We change variables using $\theta=Q x_a-\vphi$, as defined by
Eq.\rf{mag_pam}, and use the fact that in the commensurate state the
sine-Gordon nonlinearity is strong, forcing $\theta$ to fluctuate about
$0$. Thus in this phase, the dynamics is well captured by a linear
approximation in $\theta$
\bse
\begin{align}
\partial_t\theta
=& 
-\gamma\left(\tilde\epsilon_0 + 4 D_{bc}\right)\theta
+{\Gamma}m/{\chi_m} -\zeta_\vphi
\\
\partial_t m 
=&
 D_m \nabla^2 m 
- \Gamma |\psi_0|^2\left(\tilde\epsilon_0
+4D_{bc}\right)\theta + \zeta_m,
\end{align}
\ese
allowing us to straightforwardly compute the hydrodynamic correlation
functions.  Fourier transforming, we find
\begin{widetext}
\begin{equation}
\begin{pmatrix}
\theta_{{\bf k},\omega}\\ m_{{\bf k},\omega}
\end{pmatrix}  = 
{(-\omega^2+\tilde\Omega_{\bf k}^2-i \omega\tilde D_{\bf k} {k}^2)^{-1}} 
\begin{pmatrix}
-i\omega+D_m {k}^2 & {\Gamma}/{\chi_m}\\ 
-\Gamma|\psi_0|^2(\tilde{\epsilon}_{0,{\bf k}} + 4D_{bc}) &-i\omega+\gamma
 (\tilde{\epsilon}_{0,{\bf k}}+4D_{bc})
\end{pmatrix}
\begin{pmatrix}
-\zeta^\vphi_{{\bf k},\omega}\\ \zeta^m_{{\bf k},\omega}
\end{pmatrix}
\end{equation}
\end{widetext}
where $\tilde D_{\bf k} = \gamma( \tilde{\epsilon}_{0,{\bf k}} +4D_{bc})/{k}^2 
+ D_m$, $\tilde\Omega_{\bf k}=\tilde v_{\bf k} (\tilde{\epsilon}_{0,{\bf k}}+4D_{bc})^{1/2}$ 
and $\tilde v_{\bf k}=\left({\Gamma^2|\psi_0|^2}/{\chi_m}+\gamma
  D_m {k}^2\right)^{1/2}\approx {\Gamma|\psi_0|}/ {\sqrt{\chi_m}}$.

Using Gaussian noise statistics
(Eqs.~\eqref{gaussiannoise_a},\eqref{gaussiannoise_b}) to average over
$\zeta_{\vphi}$ and $\zeta_m$, we obtain \bse
\begin{align}
C_{\theta\theta}
\simeq&
\frac{2k_B T\left(\gamma\omega^2
+ D_m\tilde v_{\bf k}^2 { k}^2\right)/|\psi_0|^2}
{(\omega^2-\tilde\Omega_{\bf k}^2)^2+\omega^2  \Delta_{\bf k}^2} 
\\
C_{m m} 
\simeq&
\frac{
2k_BT\chi_m [\gamma ( \tilde{\epsilon}_{0,{\bf k}} +4D_{bc})  \tilde\Omega_{\bf k}^2 + D_m\omega^2 {k}^2]}
{(\omega^2-\tilde\Omega_{\bf k}^2)^2+\omega^2 \Delta_{\bf k}^2} 
\\
C_{\theta m} 
\simeq&
\frac{i2k_B T \Gamma \omega \Delta_{\bf k}  }{(\omega^2
-\tilde\Omega_{\bf k}^2)^2+\omega^2 \Delta_{\bf k}^2} ,
\end{align}
\ese
where $\Delta_{\bf k} \equiv \tilde D_{\bf k} k^2 \simeq 4 \gamma D_{bc}$
and we neglected terms that are subdominant in the
long-wavelength, low-frequency limit. 

We note, that as anticipated, in contrast to the incommensurate phase
of the previous subsection, here, inside the commensurate state we
find a gapful spectrum, set by the single-ion anisotropy $D_{bc}$.
This is consistent with the microscopic spin-wave analysis of
Sec.~\ref{sec2} and with experimental
observations\cite{Parshall12}.

\section{Summary}
\label{sec6}

In this work, we formulated an $S=1$ exchange model of magnetic and
structural ordering in Fe$_{1+y}$Te, a simplest parent compound of
Fe-based high-temperature superconductor.  In addition to the
exchange~\cite{Turner09}, we incorporated a competing ingredient of
single-ion orthorhombic anisotropy.  We demonstrated that this model
exhibits experimentally observed commensurate bicollinear AFM and
incommensurate spiral states, mapped out the corresponding zero
temperature phase diagram and a low energy spin-wave excitation
spectra, latter predicted to be gapped (consistent with experiments)
in the commensurate bicollinear AFM state. 

To understand the interplay between observed tetragonal-orthorhombic
structural distortion and spiral order transitions at finite
temperatures, starting with this microscopic description we derived an
effective Landau theory and used it to qualitatively map out the
temperature-doping phase diagram. Generalizing it to hydrodynamics,
that falls in the variant of the model E universality class of
Hohenberg-Halperin classification~\cite{Hohenberg77,Halperin69}, we computed the
static and dynamic structure functions $S({\bf q},\omega)$ in the PM,
commensurate bicollinear AFM and incommensurate spiral states. The
evolution of the predicted inelastic peaks with temperature and
strength of the single-ion anisotropy shows qualitative features
observed in the experiments, shifting from the incommensurate to
commensurate positions as the ordered state is approached from the
above. We hope that these predictions stimulate further systematic
experimental studies of this interesting magnetic system.

\section{Acknowledgments}

This work was supported by the NSF through DMR-1001240 (LR), MRSEC
DMR-0820579 (LR), by DOE award DE-SC0003910 (GC). We thank D. Parshall
and D. Reznik, for sharing their experimental data before publication
and discussions.

\bibliography{refs}

\end{document}